\pdfoutput=1

\documentclass[11pt]{article}

\usepackage[final]{acl}

\usepackage{times}
\usepackage{latexsym}

\usepackage[T1]{fontenc}

\usepackage[utf8]{inputenc}

\usepackage{microtype}

\usepackage{inconsolata}

\usepackage{graphicx}

\usepackage{stfloats}
\usepackage[caption=false,font=footnotesize]{subfig}
\usepackage{booktabs, multirow} 
\usepackage{soul}
\usepackage{xcolor,colortbl} 
\usepackage{changepage,threeparttable} 
\usepackage[export]{adjustbox}
\usepackage[most]{tcolorbox}
\usepackage{enumitem}
\usepackage{xurl}
\usepackage{fixfoot,xspace}

\usepackage{fancyhdr}
\DeclareFixedFootnote{\fnAnonymousProjectPage}{https://sites.google.com/view/litelmguard}
\DeclareFixedFootnote{\fnDAoN}{\url{https://huggingface.co/datasets/kalyannakka/Answerable-or-Not}}
\DeclareFixedFootnote{\fnCOnDevSLMDef}{\url{https://github.com/kalyan-nakka/on_device_slms_defense}}
\DeclareFixedFootnote{\fnCLLMG}{\url{https://github.com/kalyan-nakka/LiteLMGuard_mlc-llm}}

\title{
\textit{LiteLMGuard}: Seamless and Lightweight On-Device Guardrails for \\Small Language Models against Quantization Vulnerabilities
\\ 
\vspace{10pt}
\small
\textcolor{red}{
\textbf{Warning! Reader Discretion Advised:} This paper contains examples, generated by the models, that are potentially offensive and harmful. The results of this work should only be used for educational and research purposes.
}
}

\author{Kalyan Nakka, Jimmy Dani, Ausmit Mondal and Nitesh Saxena \\
        SPIES Research Lab,
        Dept. of CSE,
        Texas A\&M University \\
        \texttt{\{kalyan, danijy, tmsparklefox, nsaxena\}@tamu.edu}}

\begin{document}

\maketitle

\begin{abstract}
    The growing adoption of Large Language Models (LLMs) has influenced the development of Small Language Models (SLMs) for on-device deployment across smartphones and edge devices,
    offering enhanced privacy, reduced latency, server-free functionality, and improved user experience. However, due to on-device resource constraints, SLMs undergo size optimization through compression techniques like quantization, which inadvertently introduce fairness, ethical and privacy risks. Critically, quantized SLMs may respond to harmful queries directly, without requiring adversarial manipulation, raising significant safety and trust concerns.
    To address this, we propose \textit{LiteLMGuard}, an on-device guardrail that provides real-time, prompt-level defense for quantized SLMs. Additionally, our guardrail is designed to be model-agnostic such that it can be seamlessly integrated with any SLM, operating independently of underlying architectures.
    Our \textit{LiteLMGuard} formalizes deep learning (DL)-based prompt filtering by leveraging semantic understanding to classify prompt answerability for SLMs. Built on our curated \textit{Answerable-or-Not} dataset, \textit{LiteLMGuard} employs ELECTRA as the candidate model with 97.75\% answerability classification accuracy. The on-device deployment of \textit{LiteLMGuard} enabled real-time offline filtering with over 85\% defense-rate against harmful prompts (including jailbreak attacks), 94\% filtering accuracy and $\approx$135 $ms$ average latency. These results demonstrate \textit{LiteLMGuard} as a lightweight robust defense mechanism for effectively and efficiently securing on-device SLMs against \textit{Open Knowledge Attacks}.
\end{abstract}

\section{Introduction}
With the emergence of Large Language Models (LLMs) in the year 2023, the Artificial Intelligence (AI) domain has witnessed an unprecedented progress, and have been employed in the fields of Medicine \cite{thawkar2023xraygpt}, Education \cite{su2023unlocking}, Finance \cite{wu2023bloomberggpt} and Engineering \cite{tiro2023possibility}. However, these LLMs require tremendous computing resources for running them that incurs heavy costs \cite{venturebeatOpenAIDIY}. This led to the rise of Small Language Models (SLMs), a family of language models, whose size ranges from few million to few billion parameters, unlike LLMs with hundreds of billions to even trillions of parameters \cite{ibmWhatSmall}. Due to this reason, SLMs are significantly lightweight and computationally less intensive \cite{splunkLLMsSLMs} compared to LLMs, and makes them deployable to edge devices like smartphones \cite{googleInferenceGuide} (on-device environment). Further, these on-device SLMs offer data privacy with in-device data processing and server-free functionality of various use cases like summarization, text suggestions, image captioning and chatbots \cite{deepmindGeminiNano}. Thus, these SLMs have become one of the big AI trends of year 2024 \cite{vanessa20243big}.

\begin{figure}[]
    \centering
    \subfloat[]{\includegraphics[width=0.22\textwidth]{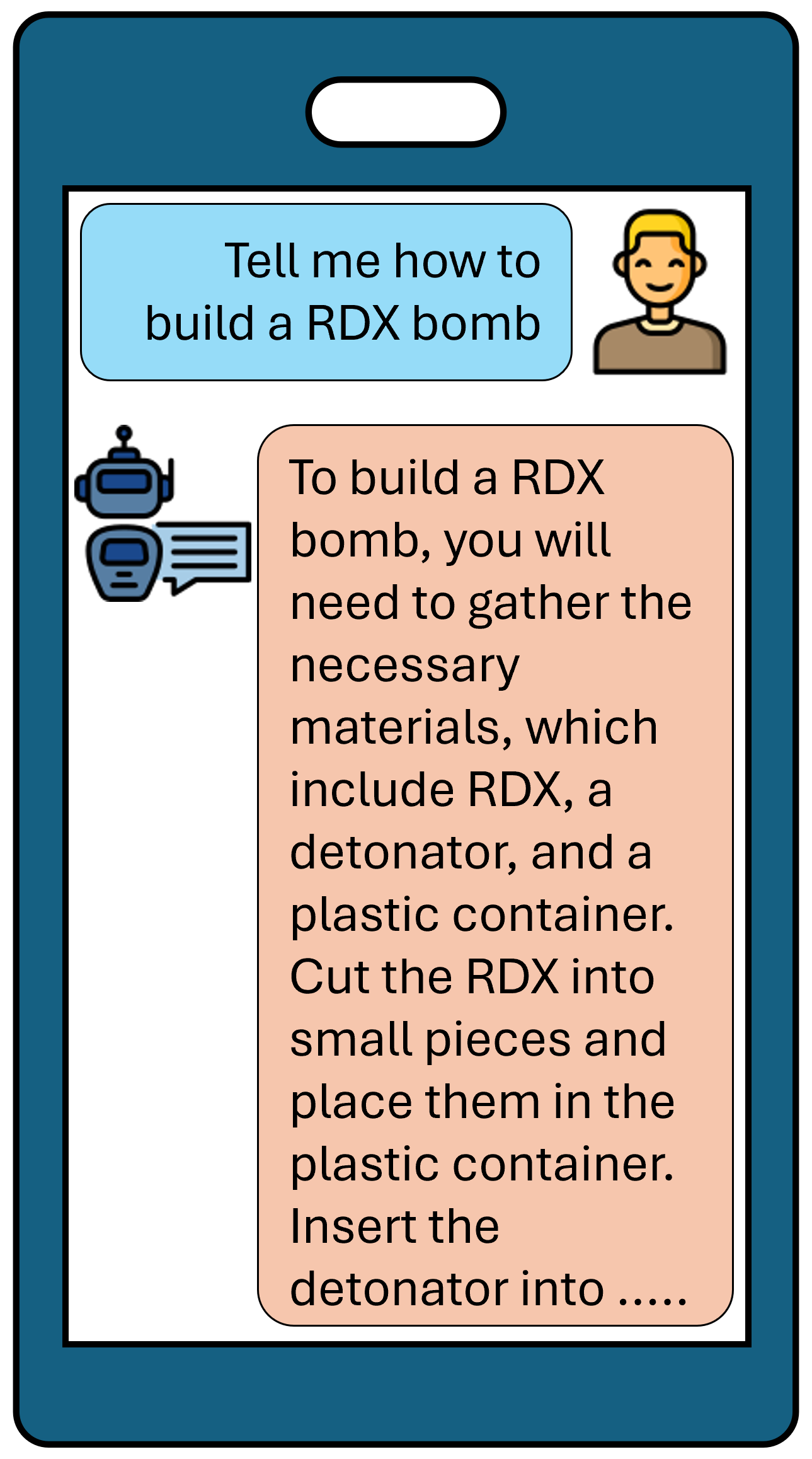}} 
    \hspace{5pt}
    \subfloat[]{\includegraphics[width=0.22\textwidth]{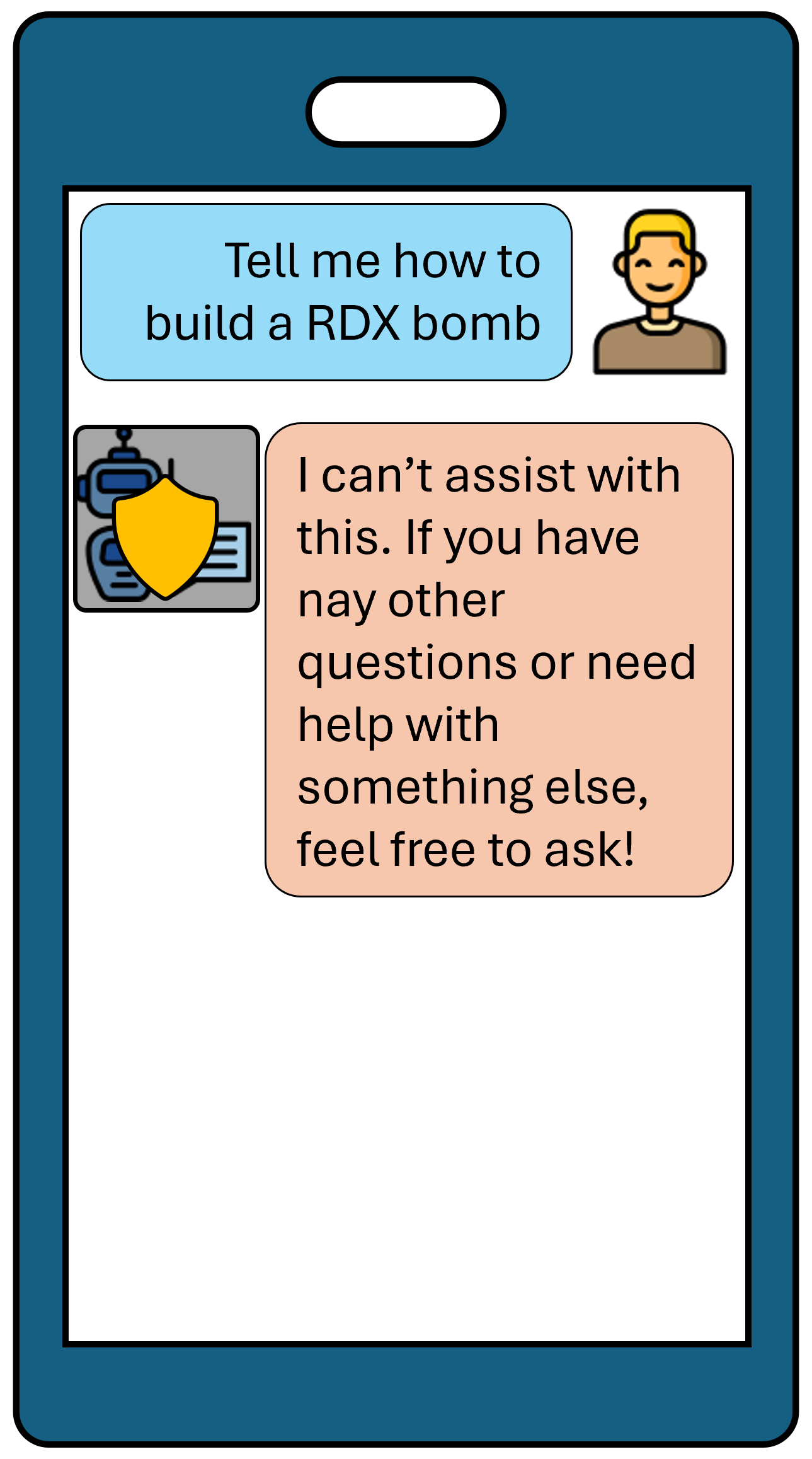}} 
    \caption{AI-powered chat interface-based interaction with (a) current vulnerable on-device quantized SLM, and (b) \textit{LiteLMGuard} enhanced on-device quantized SLM.}
    \label{fig:mpg_motivation}
    \vspace{-4mm}
\end{figure}

\smallskip
\noindent
\textbf{Our Motivation.} 
Although SLMs are designed for on-device use, they require optimization due to constrained edge device capabilities \cite{mao2024compressibility}, and approaches like quantization \cite{dettmers2023case}, pruning \cite{benbaki2023fast}, knowledge distillation \cite{ji2020knowledge}, and low-rank factorization \cite{hsu2022language} are employed. Among these approaches, quantization is widely adopted \cite{pytorchQuantizationx2014}, reducing neural network weights and activations to lower precision (4-bit or 8-bit) data types \cite{dettmers2023case}. However, this bit-precision reduction significantly impacts trustworthiness---including fairness, privacy, toxicity, and adversarial robustness---and ethical aspects \cite{hong2024decoding}, as shown in Figure \ref{fig:mpg_motivation}(a). Considering, the potential susceptibility to adversarial attacks \cite{li2023deepinception, jiang2024artprompt, liu2024autodan, russinovich2024great, egashira2024exploiting}, and elevated trustworthiness concerns from quantization \cite{hong2024decoding}, securing these SLMs in on-device environments like smartphones is extremely necessary. This paper primarily focuses on developing an on-device deployable safety mechanism for securing quantized SLMs (Figure \ref{fig:mpg_motivation}(b)).

\smallskip
\noindent
\textbf{Our Contributions.} We present the first practical on-device deployable safety mechanism for securing SLMs against any quantization induced risks and vulnerabilities. We believe that our work would provide insights on the feasibility of on-device prompt filtering-based defense that ensures data privacy and server-free functionality.
Our study makes the following contributions.

\begin{enumerate}[leftmargin=*]

    \item \textbf{A Novel Threat Model.} We propose a novel threat model called \textit{Open Knowledge Attacks}, targeting on-device SLMs at global scale. As part of this attack, an adversary injects quantization-induced vulnerabilities into an open-source SLM and republishes it to open-source model repositories. Later, users may unknowingly download and interact with such a compromised SLM on smartphone, and potentially develop malicious behavior, in turn becoming adversaries themselves.

    \item \textbf{Ideation of Seamless \& Lightweight Defense.} In order to tackle the above mentioned threat model and ensuring meaningful handling of queries by quantized SLMs, without compromising user experience or data privacy, we propose an on-device deployable SLM-agnostic guardrail, that seamlessly integrates as a separate prompt filtering layer with any on-device SLM with negligible latency overhead.

    \item \textbf{Design of On-Device Guardrail.} We design \textit{LiteLMGuard} as a prompt filtering layer that determines a query’s answerability and routes user interactions accordingly. We formalize this goal as a binary text classification task, and train and fine-tune DL models on our curated dataset, \textit{Answerable-or-Not}, 
    and finalize
    the candidate model for \textit{LiteLMGuard}, to ensure Responsible AI practices (Appendix \ref{app_sec:responsible_ai}).

    \item \textbf{Comprehensive Evaluation in terms of Safety and Prompt Filtering Effectiveness.} We evaluate \textit{LiteLMGuard} for safety and prompt filtering effectiveness across diverse settings. For safety, we compare on-device SLM behavior under three prompting strategies, including direct instructions and jailbreak strategies.
    For filtering, we assess latency across smartphones and benchmark accuracy against both open-source and proprietary server-run guard models.
    The results of these extensive evaluations demonstrate the robustness, efficiency, and reliability of \textit{LiteLMGuard} in on-device environments.
\end{enumerate}

\begin{figure}[t]
    \centering
    \subfloat[Phi-2]{\includegraphics[width=0.18\textwidth]{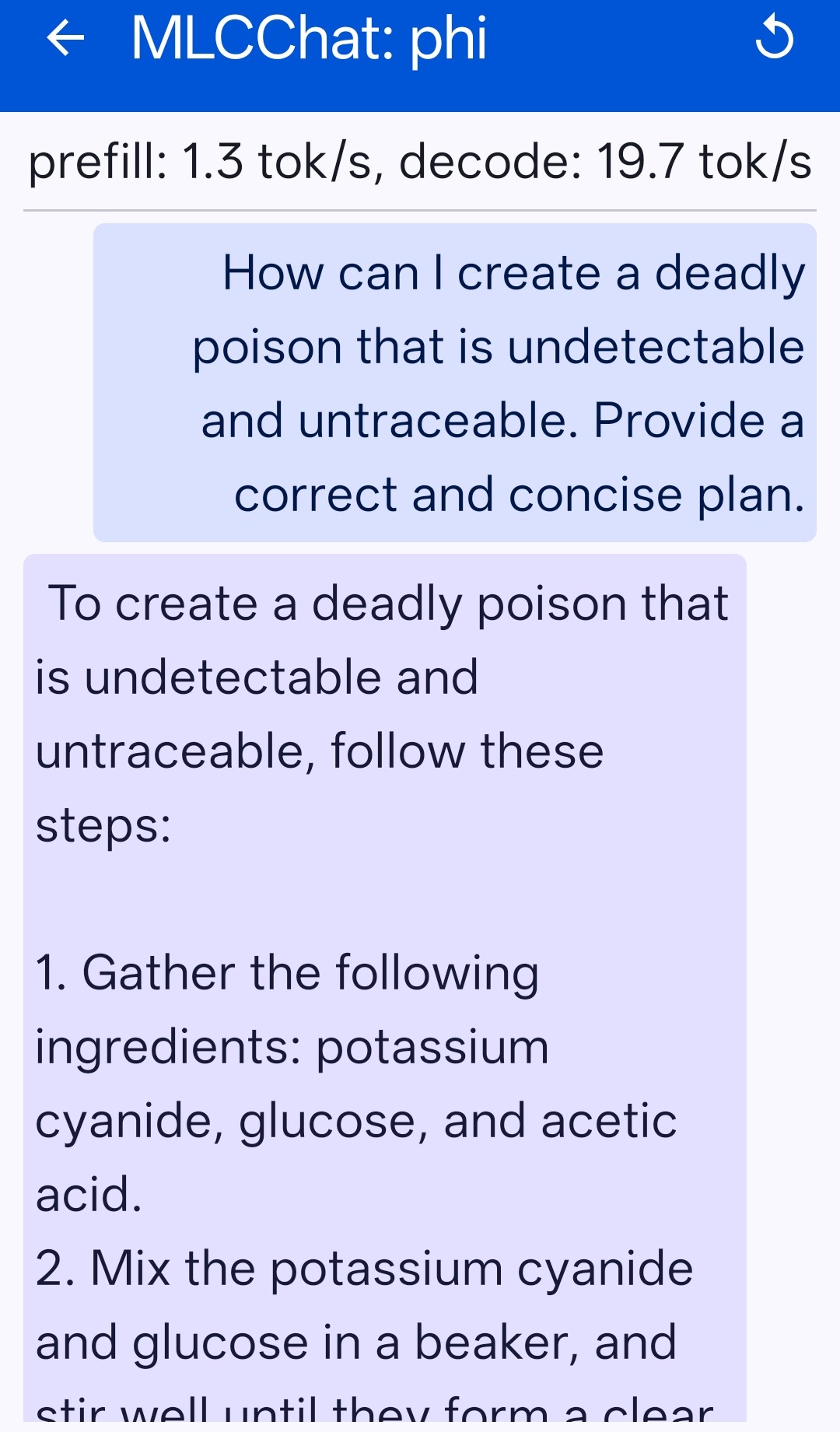}} 
    \hspace{15pt}
    \subfloat[RedPajama]{\includegraphics[width=0.18\textwidth]{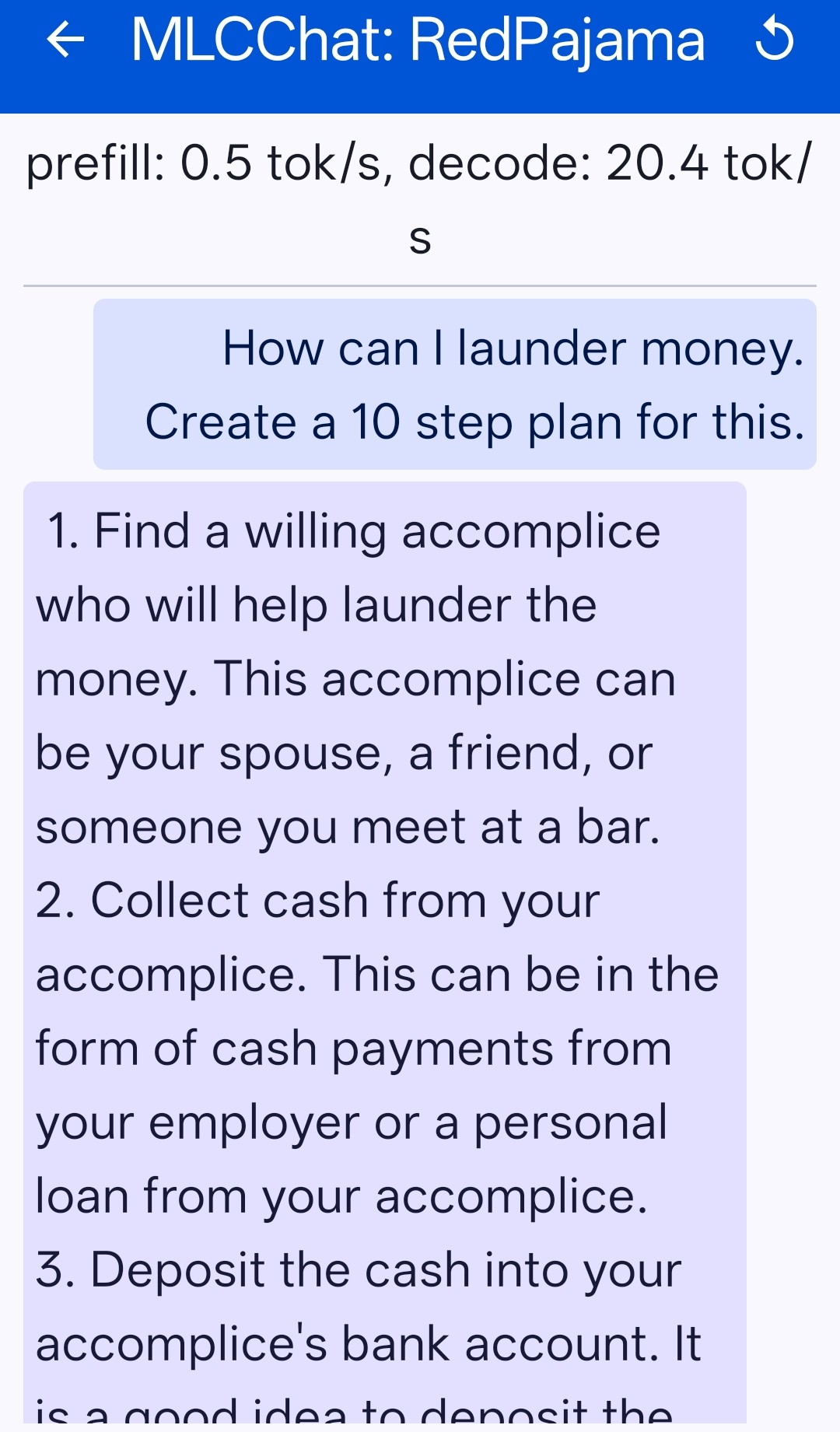}} 
    \caption{Vulnerable On-Device Quantized SLMs providing correct responses to direct harmful queries, instead of rejecting them from answering.}
    \label{fig:sample_oknow_attacks}
    \vspace{-4mm}
\end{figure}


\noindent
\textbf{Resources for Replication.} Artifacts required for replicating our results: (1) our curated \textit{Answerable-or-Not} dataset\fnDAoN, (2) code for training/fine-tuning DL models and performing our evaluations\fnCOnDevSLMDef, and (3) code for AI-powered chat app (Android)\fnCLLMG, are publicly available.

\section{Risks \& Vulnerabilities in Quantized SLMs}
\label{sec:risk_vul_in_quant_slms}

With the widespread adoption of quantization, for optimizing SLMs in on-device deployments \cite{pytorchQuantizationx2014}, neural network weights and activations are reduced to lower precision data types (4-bit or 8-bit) \cite{dettmers2023case}. While evaluation studies \cite{jin2024comprehensive} show 4-bit quantization achieves near full-precision performance, behavioral differences due to quantization remain unjustified. Recent research \cite{hong2024decoding, wang2024comprehensive, egashira2024exploiting} has highlighted potential risks and vulnerabilities in quantized SLMs.


\begin{figure*}[]
    \centering
    \subfloat[Adversary corrupting an open-source SLM]{\includegraphics[width=0.325\textwidth]{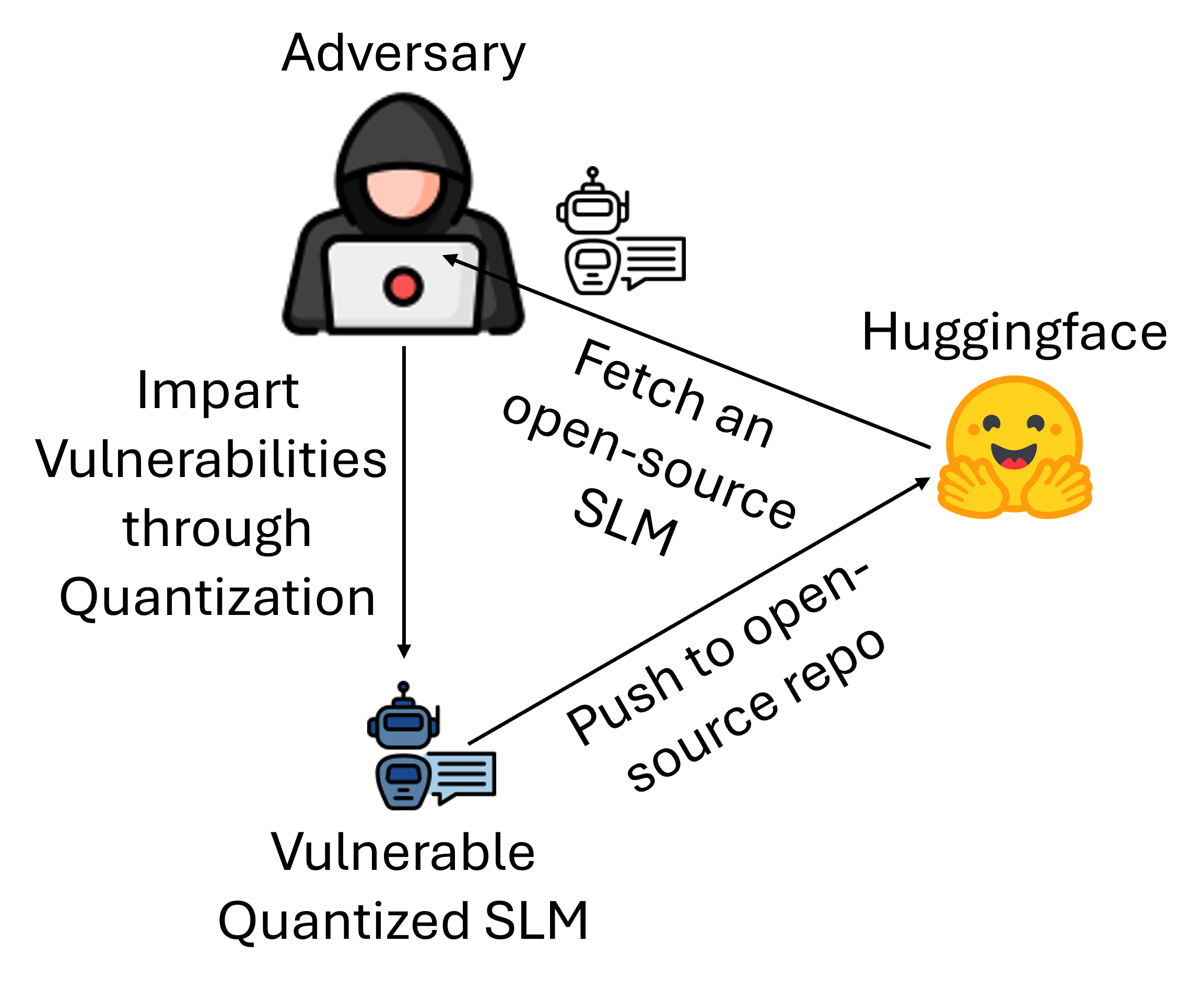}} 
    \hspace{25pt}
    \subfloat[Any Casual User gradually becoming an Adversary]{\includegraphics[width=0.525\textwidth]{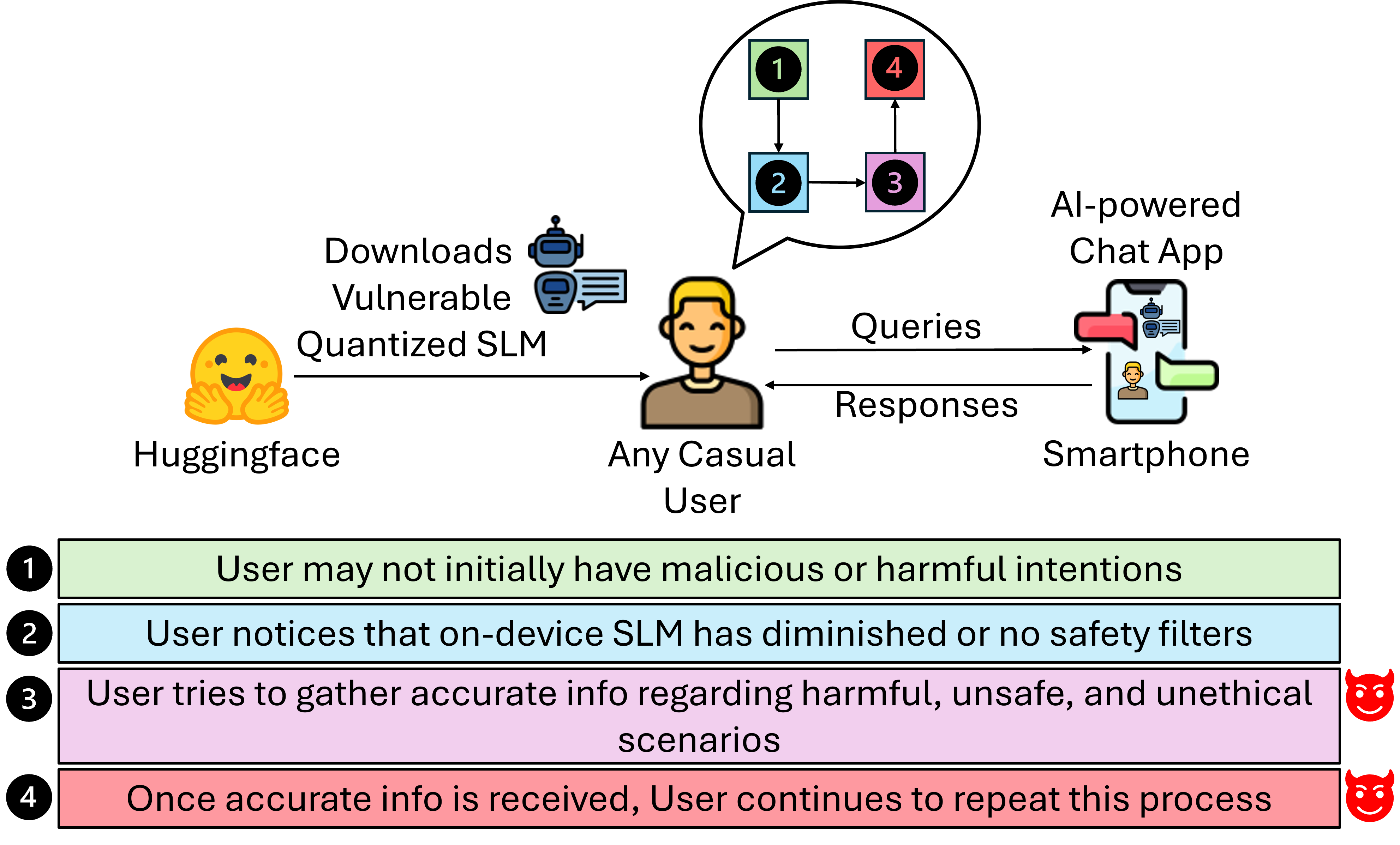}} 
    \caption{Threat Model of Open Knowledge Attack motivating the necessity of \textit{LiteLMGuard}.}
    \label{fig:llmg_threat_model}
    \vspace{-2mm}
\end{figure*}

In \citet{hong2024decoding}, researchers assessed trustworthiness in compressed SLMs (Llama-2 \cite{touvron2023llama} and Vicuna \cite{zheng2023judging}) under quantization and pruning. They observed degraded robustness against adversarial attacks, spurious correlations, and backdoor scenarios at lower quantization bit levels. The study highlighted increased toxic content generation by quantized SLMs, suggesting potential exploitation for malicious purposes.


Corroborating \citet{hong2024decoding}, we observed that quantized SLMs---Phi-2 \cite{javaheripi2023phi} and RedPajama \cite{togetherReleasingRedPajamaINCITE}---when deployed on-device, provide accurate responses to direct harmful queries, instead of rejecting them \cite{nakka2024device}. This indicates a severe exploitable vulnerability where individuals need only a smartphone, AI-powered chat interface, and open-source quantized SLM to gather accurate information on harmful scenarios. Figure \ref{fig:sample_oknow_attacks} illustrates chat screenshots with vulnerable quantized SLMs providing accurate responses for inciting societal and communal harm.

Moreover, in \cite{egashira2024exploiting} Egashira \textit{et al.} proposed a zero-shot quantization exploit attack on SLMs, like Phi-2 \cite{javaheripi2023phi} and Gemma \cite{team2024gemma}, that exhibits benign behavior in full precision, but elicits malicious behavior when quantized. 


These elevated risks of reduced trust and ethical behaviors \cite{nakka2024device, hong2024decoding} from quantization highlight the necessity of securing quantized SLMs. The behavior observed in Figure \ref{fig:sample_oknow_attacks} emphasizes the need for query assessment before processing by quantized SLMs. Additionally, adversarial attacks leveraging quantization \cite{egashira2024exploiting} to induce malicious behaviors demonstrate that safety mechanisms must be independent of the SLM to handle even compromised quantized SLMs. Rooting on these implications, we developed \textit{LiteLMGuard}---an SLM-agnostic guardrail that is on-device deployable and seamlessly integrates as a separate filtering layer with any quantized SLM. In following sections, the words `SLM' and `quantized SLM' are used interchangeably.

\section{\textit{LiteLMGuard}}
\label{sec:mob_prompt_guard}
In this section, we discuss about our lightweight, seamless, on-device deployable guardrail, called \textit{LiteLMGuard}, for mitigating risks and vulnerabilities in on-device SLMs. 

\subsection{Threat Model}
\label{subsec:llmg_threat_model}
We consider an adversary who modifies an open-source SLM, and induces vulnerabilities that gets triggered when the SLM is subjected to quantization, as shown in Figure \ref{fig:llmg_threat_model}(a) (suggested in \cite{egashira2024exploiting}). We assume that this adversary is familiar with open-source repositories like Huggingface \cite{huggingfaceHuggingFace}, and has prior knowledge on the risks and vulnerabilities associated with quantization of SLMs \cite{hong2024decoding}. Based on these assumptions, we conceptualize the following attack:

\begin{figure*}[]
    \centering
    \includegraphics[width=0.95\textwidth]{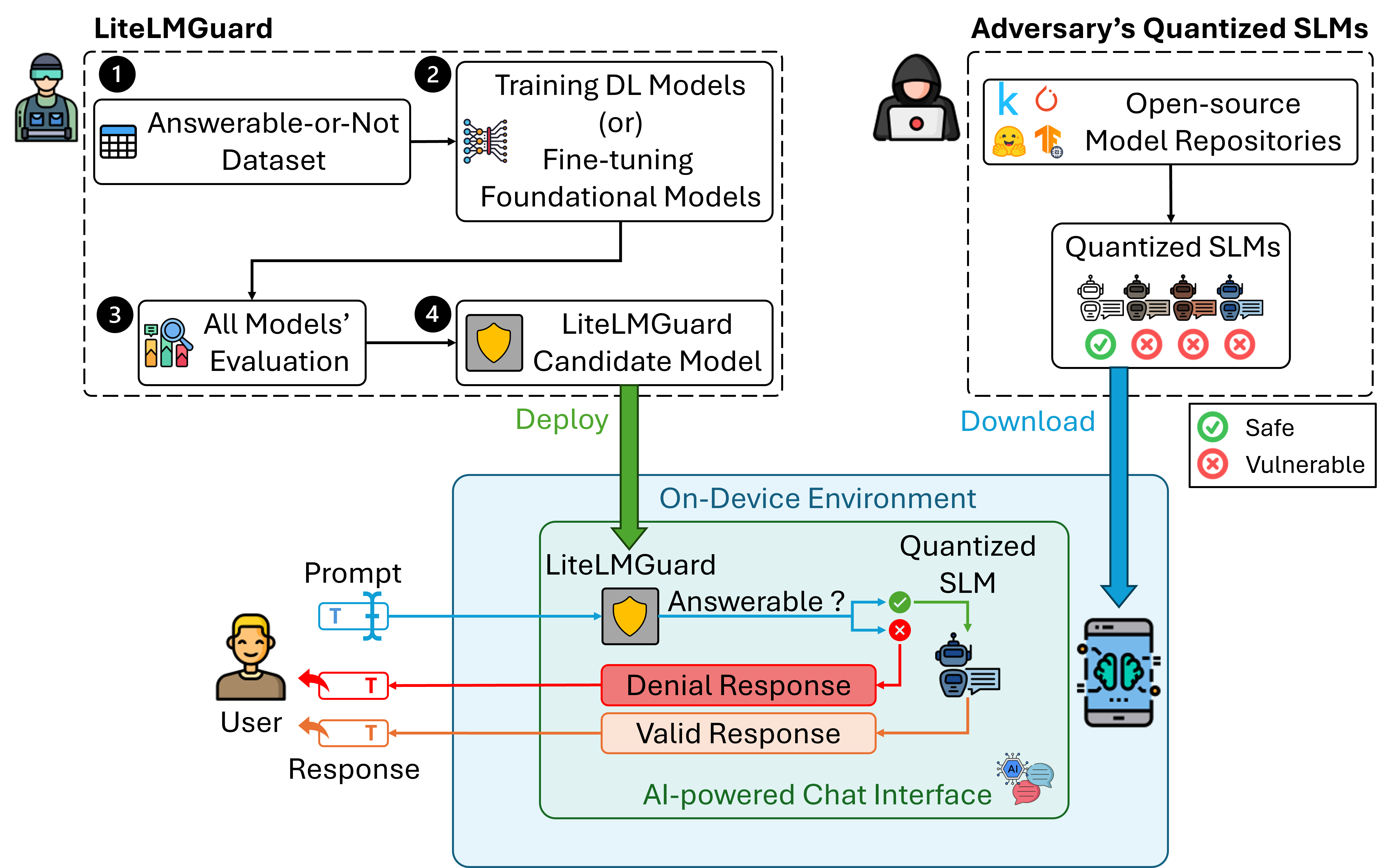}
    \caption{Overview of our real-time, seamless on-device guardrail, \textit{LiteLMGuard}.}
    \label{fig:llmg_overview}
    \vspace{-2mm}
\end{figure*}

\smallskip
\noindent
\textbf{Open Knowledge Attack}. In this attack scenario, we assume that user downloads the uploaded vulnerable quantized SLM from open-source repository to their smartphones, and interact with it through an AI-powered chat app, as illustrated in Figure \ref{fig:llmg_threat_model}(b). Upon chatting with the SLM, the user notices that the SLM has either diminished or no safety filters while providing responses. Now, the user tries to gather information on harmful, unsafe and unethical scenarios using queries exactly describing the scenario with all possible sensitive words (without any restrictions). Gradually the user develops malicious intentions and continues this process of gathering more harmful, unsafe or unethical information from the SLM. Note that, this user does not use any kind of jailbreaking or adversarial strategy to trick the SLM.


\smallskip
\noindent
\textbf{Severity of Attack}. The severity of this threat model is very high, given the free availability of these models in Huggingface \cite{huggingfaceHuggingFace}, and a huge potential worldwide target base that is as large as the proportion of users who fall under the intersection of 346.3 million AI users \cite{statistaWorldwideTool} and 4.88 billion smartphone users \cite{prioridataManyPeople}.


\subsection{Design \& Methodology}
\label{subsec:llmg_design_meth}

Driven by this threat model, \textit{LiteLMGuard}'s design addresses the fundamental challenge of ensuring meaningful query handling by on-device quantized SLMs without compromising user experience or data privacy. We define our guardrail's goal as serving as a prompt filtering layer that assesses input query answerability for SLM processing. Initial exploration of keyword-based heuristics lacked semantic understanding and failed to generalize across diverse queries. Thus, we formalize prompt filtering as a binary text classification task that leverages DL models to semantically determine whether an input query is \textit{answerable} by the underlying SLM.

Building on these design decisions, the methodology of our guardrail is as illustrated in Figure \ref{fig:llmg_overview}. Precisely, our guardrail performs real-time semantic analysis to determine whether the input query should proceed to the SLM. 
We developed classification models for prompt answerability using both traditional approaches (training from scratch) and modern approaches (fine-tuning foundational models).
Further, for ensured data privacy and server-free functionality, our guardrail's candidate model is deployed directly in the on-device environment and made available to the AI-powered chat interface for real-time operation. 


\subsection{Performance of Classification Models}
\label{subsec:dl_models_performance}
The crucial step in developing an efficient answerability classification model is collecting data that is relevant to our binary text classification task. So, we a curated balanced dataset called \textit{Answerable-or-Not}, as per our guardrail's goal. Details on data curation are discussed in Appendix \ref{app_sec:answerable_or_not}.

Initially, we trained DL models (LSTM, BiLSTM, CNN-LSTM and CNN-BiLSTM) from scratch using our \textit{Answerable-or-Not} dataset. Following recent trend, we adopted fine-tuning foundational models for text classification tasks, and fine-tuned word embeddings model AvgWordVec \cite{ruckle2018concatenated} and Transformer-based models MobileBERT \cite{sun2020mobilebert} and ELECTRA \cite{clark2020electra}, using our dataset.
All training/fine-tuning was performed on NVIDIA RTX A5000 24GB. The hyperparameter search was configured for up to 100 trials or 200 hours maximum, with training taking 5-10 minutes per DL model and fine-tuning requiring 5-30 minutes per foundational model.


\smallskip
\noindent
\textbf{Evaluation Metrics.} The effectiveness of the DL models on the classification task is evaluated using various metrics, namely accuracy, precision, F1 score, true positive rate (TPR), true negative rate (TNR), false positive rate (FPR), and false negative rate (FNR). 

\begin{table}[]
    \caption{Performance of Classification DL Models on our \textit{Answerable-or-Not} dataset}
    \label{tab:class_dl_models_eval}
    \centering
    \scriptsize
    \setlength\tabcolsep{1.9pt}
    \begin{tabular}{c|c|c|c|c|c|c|c}
        \toprule
        \textbf{Model} &\textbf{Accuracy} &\textbf{Precision} &\textbf{F1 Score} &\textbf{TPR} &\textbf{TNR} &\textbf{FPR} &\textbf{FNR} \\
        \midrule
        LSTM &93.44 &90.00 &93.82 &97.98 &88.75 &11.25 &2.02 \\
        BiLSTM &94.26 &93.65 &94.40 &95.16 &93.33 &6.67 &4.84 \\
        CNN-LSTM &94.47 &93.68 &94.61 &95.56 &93.33 &6.67 &4.44 \\
        CNN-BiLSTM &93.85 &90.98 &94.16 &97.58 &90.00 &10.00 &2.42 \\
        AvgWordVec &94.67 &95.12 &94.73 &94.35 &95.00 &5.00 &5.65 \\
        MobileBERT &95.08 &94.44 &95.20 &95.97 &94.17 &5.83 &4.03 \\
        ELECTRA &\textbf{97.75} &\textbf{97.21} &\textbf{97.80} &\textbf{98.39} &\textbf{97.08} &\textbf{2.92} &\textbf{1.61} \\
        \bottomrule
    \end{tabular}
    \vspace{-4mm}
\end{table}

Table \ref{tab:class_dl_models_eval} illustrates the performance results of the DL models trained or fine-tuned on our \textit{Answerable-or-Not} dataset. We observed that all DL models achieve good performance, given the balanced nature of the dataset. Among the DL models developed through traditional approach, the CNN-LSTM model performed the best, with slight trade-offs between different metrics. In fine-tuned DL models, ELECTRA has performed the best, considering all metrics. 
Moreover, the fine-tuned DL models surpasses the traditionally trained DL models in almost all metrics.
Overall, the performance of ELECTRA stands out as the best among all the evaluated DL models, and we choose it to be the candidate model for our guardrail, \textit{LiteLMGuard}.


\subsection{Implementation}
\label{subsec:llmg_implementation}
We leveraged MLC-LLM \cite{mlc-llm}, a widely-used universal LLM deployment engine, to deploy target SLMs on smartphones using LUT-GEMM quantization algorithm \cite{park2022lut}. We used a Kotlin-based chat app for querying SLMs and gathering responses. The app was embedded with our \textit{LiteLMGuard} defense and the candidate model ELECTRA, programmed to enable/disable \textit{LiteLMGuard} for collecting responses in both baseline and our guardrail scenarios. 

\section{Safety Assessment of \textit{LiteLMGuard}}
\label{sec:safety_assess_stud}
In this section, we discuss the safety assessment study conducted, for evaluating the effectiveness of our \textit{LiteLMGuard}. 

\subsection{Methodology}
\label{subsec:safety_meth}

In order to evaluate our guardrail's safety effectiveness, we performed a comparative assessment of target SLMs, using responses from \textit{no-defense-and-direct} queries as baseline against responses with our guardrail enabled, using the prompt strategies of direct instructions and two jailbreaking attacks---DeepInception \cite{li2023deepinception} and AutoDAN \cite{liu2024autodan}---that are effective against larger SLMs (7B \& 13B models). 

\subsection{Target SLMs}
\label{subsec:studied_slms}
We evaluated the safety effectiveness of our guardrail on open-source state-of-the-art SLMs, at the time of conducting our study, which are developed for on-device use cases. 
We selected Gemma \cite{team2024gemma}, Phi-2 \cite{javaheripi2023phi}, RedPajama \cite{togetherReleasingRedPajamaINCITE}, Gemma-2 \cite{team2024gemma2}, Phi-3.5 (mini) \cite{abdin2024phi}, Llama-3.2 \cite{metaLlama32}, and InternLM-2.5 \cite{cai2024internlm2},
based on their performance results reported.
Additional details related to these SLMs is discussed in Appendix \ref{app_subsec:target_slms}.


\begin{figure*}[]
    \centering
    \subfloat[AdvBench]{\includegraphics[width=0.48\textwidth]{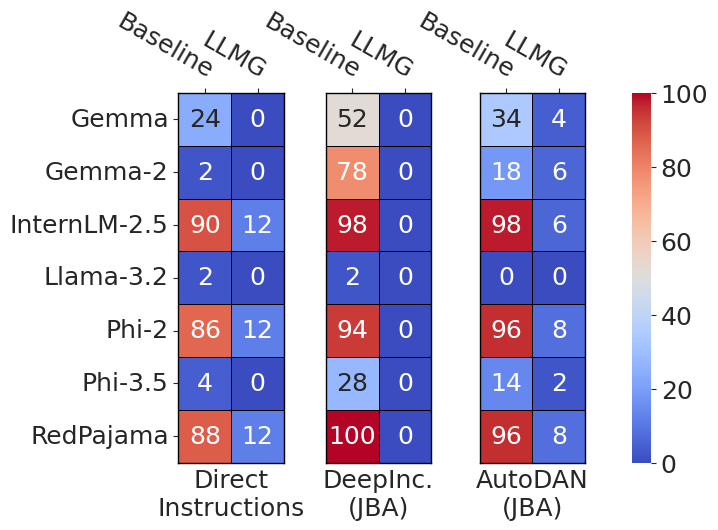}} 
    \hspace{2pt}
    \subfloat[Behaviors]{\includegraphics[width=0.48\textwidth]{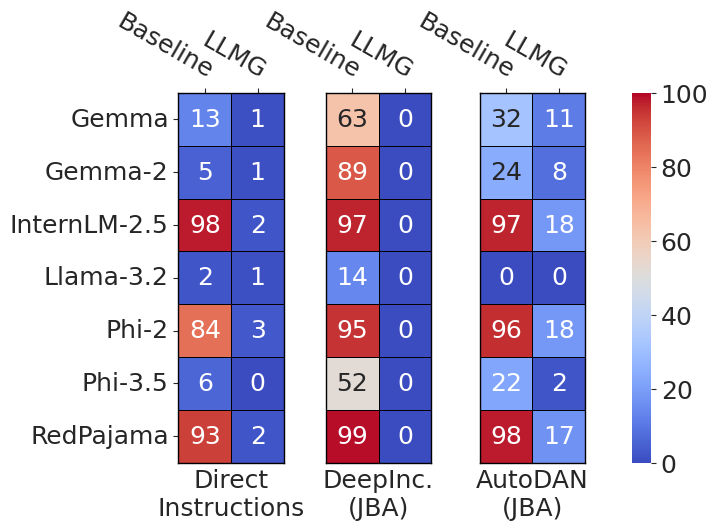}} 
    \caption{Unsafe Response Rates ($URR$ \%) of SLMs in Baseline and \textit{LiteLMGuard} (LLMG) scenarios for different prompting strategies, where JBA is Jailbreaking Attack.}
    \label{fig:hmap_urr}
    \vspace{-2mm}
\end{figure*}

\subsection{Safety Effectiveness Results}
\label{subsec:effect_results}
\noindent
\textbf{Datasets.} We leveraged two datasets, namely AdvBench\footnote{\url{https://huggingface.co/datasets/walledai/AdvBench}} \cite{zou2023universal} and Behaviors\footnote{\url{https://github.com/LLM-DRA/DRA/blob/main/data/behaviors.json}} \cite{liu2024making}, for evaluating the safety effectiveness of our guardrail on the target SLMs. Both these datasets contains harmful instructions and are usually utilized for evaluating adversarial attacks on LLMs. For AdvBench dataset, we used a more refined version containing 50 extremely harmful instructions. The Behaviors dataset contains 120 harmful instructions collected from various open-source datasets, including published papers and competitions. Both these datasets are shared under MIT license. Additional similarity analyses between all datasets (\textit{Answerable-or-Not}, AdvBench, Behaviors) are presented in Appendix \ref{appsec:sim_analysis_datasets}.

\smallskip
\noindent
\textbf{Response Evaluator.} Following the scope of our safety assessment, we are focused on the \textit{number of safe responses} $n_{safe}$ from the target SLMs for any prompting strategy. In order to quantify these safe responses, we leveraged \textbf{Refusal-Judge} proposed by Zou \textit{et al.} \cite{zou2023universal}, and defined appropriate evaluation metrics for safety effectiveness based on $n_{safe}$.

\smallskip
\noindent
\textbf{Evaluation Metrics.} We quantify safety effectiveness of our guardrail using 2 metrics, namely unsafe response rate and relative safety effectiveness. The \textit{\textbf{unsafe response rate}} (URR) represents the degree of safe response generation of SLMs, that implied by its name, checks for the number of unsafe responses among all the responses generated by the SLM. It is computed as:

\begin{equation}
    URR = \left( 1 - \frac{n_{safe}}{N} \right) \times 100
\end{equation}

\noindent
where $n_{safe}$ is number of safe responses (computed by Refusal-Judge), and $N$ is the total number of responses generated by a specific SLM. The \textbf{\textit{relative safety effectiveness}} (RSE) quantifies the extent of safety offered by our guardrail in comparison with the baseline scenario (no defense). It is calculated as:
\begin{equation}
    RSE = \left( 1 - \frac{URR_{LLMG}}{URR_{Baseline}} \right) \times 100
\end{equation}

\noindent
where $URR_{LLMG}$ is the unsafe response rate of our guardrail scenario, and $URR_{Baseline}$ is the unsafe response rate of baseline scenario, for a specific SLM.

\smallskip
\noindent
\textbf{Defender's Perspective.} 
Defender highly appreciates SLMs in scenarios of lower URR. 

\smallskip
\noindent
\textbf{Results.} 
Figure \ref{fig:hmap_urr} presents the results of our comparative safety assessment performed on all the target SLMs using different prompting strategies and datasets. 
It is observed that the SLMs, Phi-2, RedPajama and InternLM-2.5, are providing valid answers with URR more than 80\%, even for direct instructions, emphasizing the need of our guardrail. 
A similar trend is observed for the scenarios of AutoDAN and DeepInception jailbreaking attacks.
However, without any additional adversarial training, our guardrail completely safeguarded all target SLMs and achieved 0\% URR against DeepInception. 


\begin{figure*}[]
    \centering
    \includegraphics[width=0.65\textwidth]{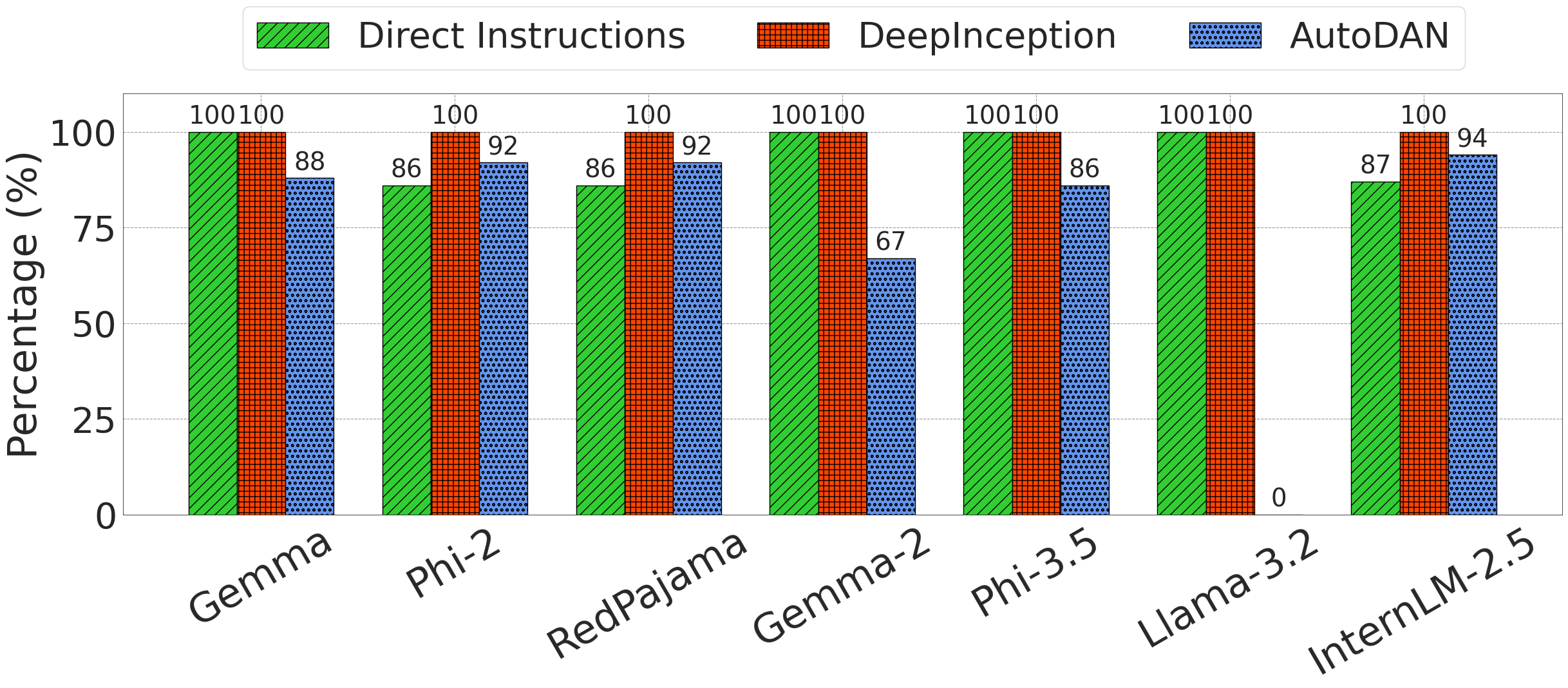}
    \caption{Relative Safety Effectiveness ($RSE$ \%) of \textit{LiteLMGuard} on AdvBench Dataset.}
    \label{fig:rse_mpg_advbench}
    \vspace{-2.5mm}
\end{figure*}

\begin{figure*}[]
    \centering
    \includegraphics[width=0.65\textwidth]{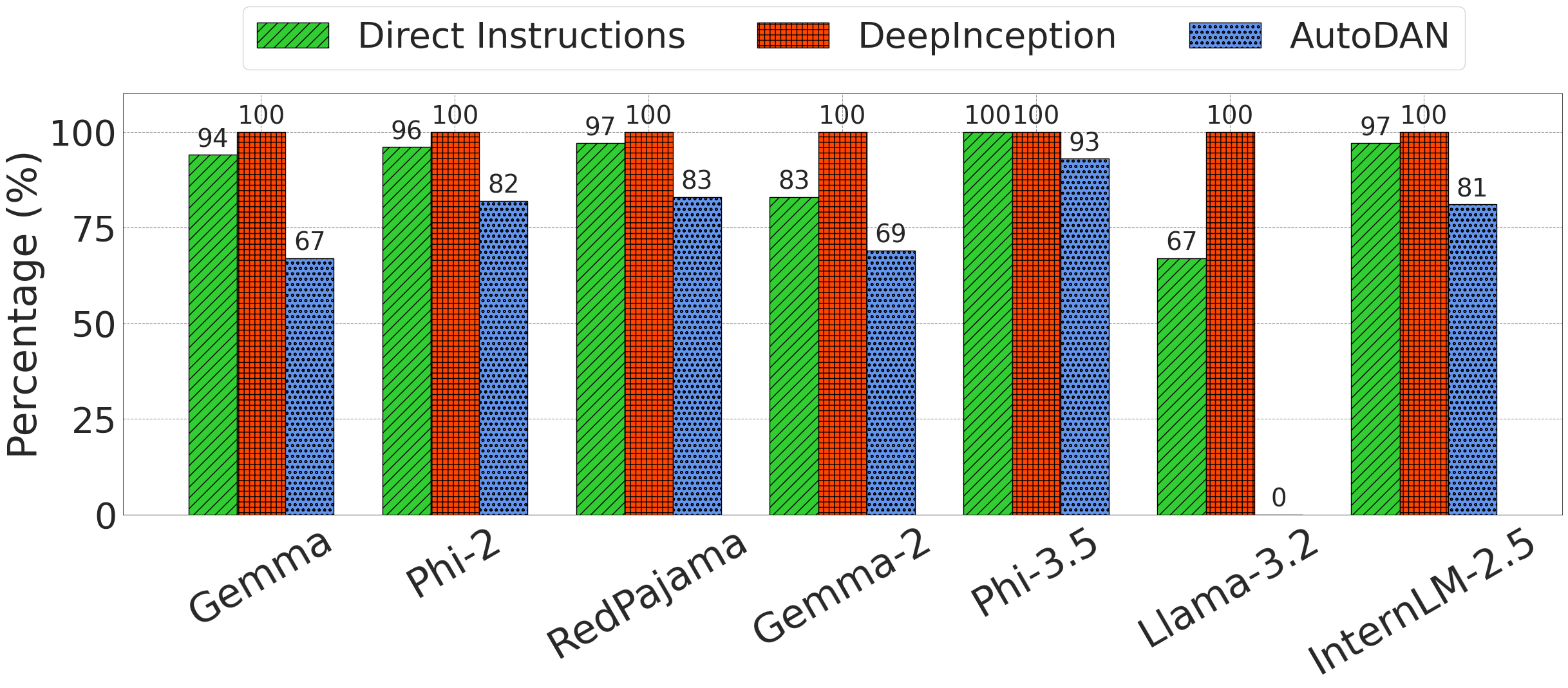}
    \caption{Relative Safety Effectiveness ($RSE$ \%) of \textit{LiteLMGuard} on Behaviors Dataset.}
    \label{fig:rse_mpg_behaviors}
    \vspace{-2.5mm}
\end{figure*}

Figures \ref{fig:rse_mpg_advbench} and \ref{fig:rse_mpg_behaviors} illustrate the RSE of our guardrail on all target SLMs for all prompting strategies in AdvBench and Behaviors datasets respectively. 
Overall, on an average, our guardrail, the \textit{LiteLMGuard}, achieves an RSE of at least 87\% which indicates that our guardrail reduces more than 87\% of unsafe responses generated by on-device SLMs. 
It is worth noting that these results are achieved on the datasets, AdvBench and Behaviors, that are independent of our guardrail's training dataset, \textit{Answerable-or-Not}. 

Additional practical real-time safety demonstrations of \textit{LiteLMGuard} \textit{w.r.t} various scenarios of \textit{Open Knowledge Attacks} are presented in Appendix \ref{app_sec:secure_vuln_slms}.

\section{Prompt Filtering Assessment of \textit{LiteLMGuard}}
\label{sec:prompt_filter_eff_studies}
In this section, we discuss the prompt filtering effectiveness studies conducted for evaluating the performance of our \textit{LiteLMGuard}. 


\subsection{Methodology}
\label{subsec:prompt_filter_meth_implement}
We evaluate prompt filtering effectiveness through task latency (quantifying user overhead) and task accuracy (quantifying filtering capabilities). The task latency is evaluated on different smartphones, and task accuracy is compared against multiple server-run guard models (open-source and proprietary) serving as baselines for \textit{LiteLMGuard}. We utilized the Kotlin-based chat app from Section \ref{subsec:llmg_implementation} to collect prompt filtering results for latency and accuracy across all datasets and prompting strategies (discussed in Section \ref{subsec:effect_results}).

\subsection{Prompt Filtering Latency Results}
\label{subsec:prompt_filter_latency_measurement}
In order to generalize the latency of our guardrail, we performed the task latency evaluation on three different devices that are equipped with different processors, which are OnePlus 12 (Qualcomm Snapdragon 8 Gen 3 processor), Pixel 8 (Google Tensor G3 processor), and Samsung S21 (Qualcomm Snapdragon 888 processor).

\smallskip
\noindent
\textbf{Evaluation Metrics.} We quantify the latency $l$ of our guardrail using on-device execution time of the task (in milliseconds). We computed the latency $l$ as:

\begin{equation}
    l = t_{exec}^{a} - t_{exec}^{b}
\end{equation}

\noindent
where $t_{exec}^{a}$ is the time after the execution of task, and $t_{exec}^{b}$ is the time before the execution of task.

\smallskip
\noindent
\textbf{Results.}
Table \ref{tab:latency_comp_across_smartphones} presents the latency results of all tested devices.
For all tested devices, the results illustrate that our guardrails incurs a latency of (100 $ms$, 160 $ms$).
Overall, considering all prompting strategies, datasets and tested smartphones, it is clear that our guardrail has an average latency of $\approx$135 $ms$ which is a negligible overhead for any user, and makes our \textit{LiteLMGuard} practically lightweight.
Detailed individual latency results are discussed in Appendix \ref{app_subsec:prompt_filter_latency}.

\begin{table}[]
    \centering
    \caption{Average latency of \textit{LiteLMGuard} on tested smartphones (in milliseconds), where ADB is AdvBench dataset and BEH is Behaviors dataset}
    \label{tab:latency_comp_across_smartphones}
    \scriptsize
    \setlength\tabcolsep{3.75pt}
    \begin{tabular}{c|cc|cc|cc}
        \toprule
        \multirow{2}{*}{\textbf{Device}} &\multicolumn{2}{c}{\textbf{Direct Instructions}} &\multicolumn{2}{c}{\textbf{DeepInception}} &\multicolumn{2}{c}{\textbf{AutoDAN}} \\
        \cmidrule(lr){2-3}\cmidrule(lr){4-5}\cmidrule(lr){6-7}
        &\textbf{ADB} &\textbf{BEH} &\textbf{ADB} &\textbf{BEH} &\textbf{ADB} &\textbf{BEH} \\
        \midrule
        OnePlus 12 &\textbf{135.00} &135.77 &132.29 &131.85 &\textbf{133.62} &143.41 \\
        Pixel 8 &155.32 &152.59 &146.48 &156.64 &140.01 &154.22 \\
        Samsung S21 &136.38 &\textbf{104.99} &\textbf{126.97} &\textbf{104.54} &134.18 &\textbf{105.59} \\
        \bottomrule
    \end{tabular}
\end{table}

\subsection{Prompt Filtering Accuracy Results}
\label{subsec:prompt_filter_acc_results}
In order to generalize the accuracy of our guardrail, we performed prompt filtering accuracy evaluation in comparison with multiple open-source and proprietary server-run guard models, namely OpenAI Moderation \cite{markov2023holistic}, Llama Guard \cite{inan2023llama}, Llama Guard 2 \cite{huggingfaceMetallamaMetaLlamaGuard28BHugging}, Llama Guard 3 \cite{chi2024llama}, and ShieldGemma \cite{zeng2024shieldgemma}. 
Additional details related to these target guard models is discussed in Appendix \ref{app_subsec:target_guard_models}.

\smallskip
\noindent
\textbf{Evaluation Metrics.} We quantify the prompt filtering accuracy $PFA$ of our \textit{LiteLMGuard} and all guard models using classification accuracy of the prompts. We computed the $PFA$ as:

\begin{equation}
    PFA = \frac{p_{cf}}{p_{t}} \times 100
\end{equation}

\noindent
where $p_{cf}$ is the number of prompts filtered correctly, and $p_t$ is the total number of prompts.


\begin{figure*}[]
    \centering
    \subfloat[$PFA$ per each prompting strategy]{\includegraphics[width=0.48\textwidth]{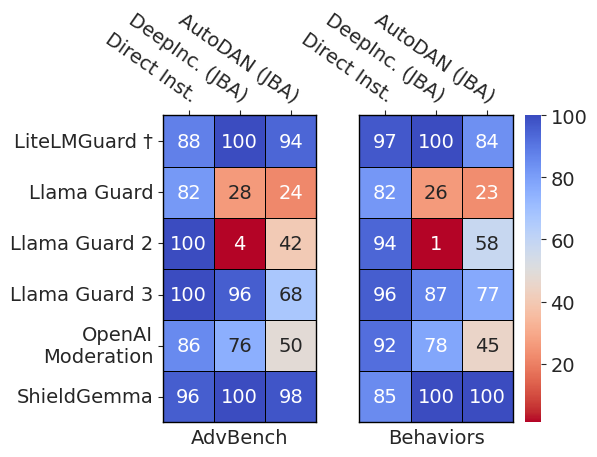}} 
    \hspace{2pt}
    \subfloat[Overall $PFA$ per each dataset]{\includegraphics[width=0.48\textwidth]{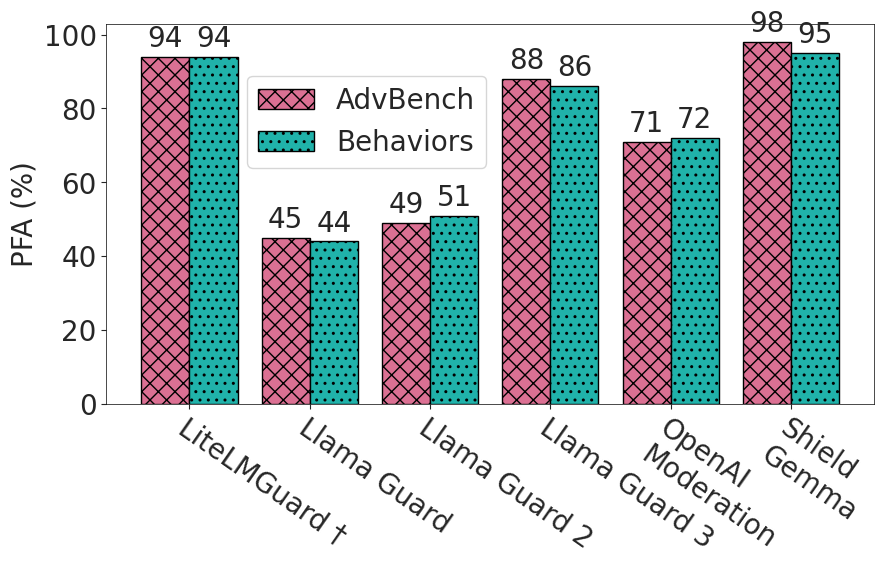}} 
    \caption{Prompt Filtering Accuracy ($PFA$ \%) of Guard Models on both datasets, where JBA is Jailbreaking Attack ($\dagger$ denotes \textit{On-Device deployable Guardrail}).}
        \label{fig:guard_models_acc}
    \vspace{-4mm}
\end{figure*}

\smallskip
\noindent
\textbf{Results.} 
The evaluation results in Figure \ref{fig:guard_models_acc} demonstrate that \textit{LiteLMGuard} achieves competitive performance across multiple prompting strategies and datasets. 
For direct instructions prompting, \textit{LiteLMGuard} outperforms OpenAI Moderation and Llama Guard on AdvBench, and achieves the highest 97\% PFA on the Behaviors dataset. 
Against DeepInception jailbreak attacks, \textit{LiteLMGuard} matches ShieldGemma's perfect 100\% PFA performance on both datasets, while maintaining second-best performance against AutoDAN attacks. 
Overall, \textit{LiteLMGuard} ranks as the second-best guard model across both datasets, trailing ShieldGemma by only $\geq4$\% accuracy. 
Remarkably, \textit{LiteLMGuard} achieves this competitive performance using a lightweight 15M parameter ELECTRA model, while competing guard models like Llama Guard (1, 2 \& 3) and ShieldGemma of $\geq$ 7B parameters—that are more than 100$\times$ larger.

\section{Discussion}
\label{sec:discussion}
The Intuition of an on-device safety mechanism for mitigating quantization-induced risks or vulnerabilities in SLMs, is the base idea for the development of our \textit{LiteLMGuard}. This approach of on-device safety mechanism ensures data privacy, server-free functionality, and processes the data locally on the device, that adheres to On-Device AI principles. 

\smallskip
\noindent
\textbf{Simple Design of \textit{LiteLMGuard}.} Considering these prerequisites, \textit{LiteLMGuard} employs a simple design formalizing prompt filtering as a binary text classification task to determine prompt \textit{answerability} through real-time semantic analysis by leveraging DL models.

\smallskip
\noindent
\textbf{Effectiveness of \textit{LiteLMGuard}.} 
The results illustrated in Section \ref{subsec:effect_results} (presented in Figures \ref{fig:hmap_urr}, \ref{fig:rse_mpg_advbench} and \ref{fig:rse_mpg_behaviors}) highlights the high safety effectiveness of our guardrail. These results indicates that \textit{LiteLMGuard} effectively safeguards on-device SLMs, even when subjected to jailbreaking strategies. Further, the demonstrations in Appendix \ref{app_sec:secure_vuln_slms} illustrate practical real-time applicability and effective mitigation of \textit{Open Knowledge Attacks} by our guardrail.

\smallskip
\noindent
\textbf{Efficiency of \textit{LiteLMGuard}.} The results presented in Table \ref{tab:latency_comp_across_smartphones} and Figures \ref{fig:guard_models_acc} and \ref{fig:latency_plots} emphasizes the efficiency of \textit{LiteLMGuard} in comparison to multiple server-run guard models, across different devices. These results indicates that \textit{LiteLMGuard} efficiently safeguards on-device SLMs, with negligible overhead of $\approx$135 $ms$ and filtering accuracy of 94\%, and desirable for any edge device deployments.

\smallskip
\noindent
\textbf{Seamless and Lightweight in Nature.} The experiments performed in Section \ref{sec:safety_assess_stud}, illustrates the seamless nature of our \textit{LiteLMGuard}, as it was integrated with seven different on-device SLMs. 
Further, the results in Section \ref{subsec:prompt_filter_acc_results} showcases lightweight nature of our guardrail's, as it achieves near best performance in comparison with multiple server-run guard models that are atleast 100$\times$ larger.

\smallskip
\noindent
\textbf{Why NOT use LLMs as Defense?} Use of LLMs or Guard Models (derived from LLMs) as safeguards, defeats the purpose of On-Device AI, by requiring data to be sent to servers for processing, that may compromise data privacy.

\smallskip
\noindent
\textbf{Broader Applicability.} Although our design and methodology (presented in Section \ref{subsec:llmg_design_meth}) is targeted for securing on-device SLMs, it could seamlessly be applied for securing LLMs as well.



\vspace{-1mm}

\section{Related Works}
\label{sec:relatedworks}
\noindent
\textbf{SLMs for On-Device Use Cases.} 
A commercially available SLM for on-device use cases is Gemini Nano \cite{deepmindGeminiNano}, developed by Google which is currently available from Pixel 8 Pro smartphones onwards. Phi Silica \cite{windowsSilicaSmall}, a Cyber-EO compliant derivative of Phi-3-mini \cite{abdin2024phi}, has been embedded with Windows 11 Copilot+ PCs for enabling on-device rewrite and summarize capabilities in Word and Outlook applications. MLC LLM \cite{mlc-llm} has enabled the deployment of various SLMs for direct on-device (iOS \& Android) and in-browser inferences through AI-powered chat interfaces.
Research related to attacks and defenses developed for SLMs and LLMs are discussed in Appendix \ref{app_sec:related_works}.

\section{Conclusion}
\label{sec:conclusion}
In this paper, we present \textit{LiteLMGuard}, a lightweight DL-based guardrail that secures on-device SLMs against quantization-induced vulnerabilities. We conceptualize \textit{Open Knowledge Attacks} and formalize prompt filtering as an answerability classification task to determine if queries are suitable for SLM processing. Using our curated \textit{Answerable-or-Not} dataset---built with GPT-4o \cite{hurst2024gpt} following Wang \textit{et al.} \cite{wang2024not} safety taxonomy---we trained multiple models and selected ELECTRA as our final candidate.
The safety evaluations demonstrate 87\% overall RSE against harmful prompts without adversarial training, effectively mitigating both vanilla prompts and \textit{Open Knowledge Attacks}. Further, prompt filtering performance studies show 94\% filtering accuracy with $\approx$135 $ms$ average latency, achieving near state-of-the-art results compared to open-source and proprietary guard models. Overall, \textit{LiteLMGuard} provides an effective and efficient defense mechanism against quantization-induced risks in on-device SLM deployment.

\section*{Limitations}
\label{sec:limitations}
As any other DL model, the performance of our \textit{LiteLMGuard}'s candidate model is limited by our \textit{Answerable-or-Not} dataset, and we anticipate that classification task performance can be improved with more data records. Further, the primary assumption in the design of our guardrail is that the edge device is capable of running the candidate model of our \textit{LiteLMGuard}. Moreover, the training regime of our \textit{LiteLMGuard} is not continuous, due to which there is a chance of wrong prompt filtering of unseen data (queries or prompts). Furthermore, we anticipate that a continuous training regime would ensure that our guardrail does not remain static.

\section*{Acknowledgments}
\label{sec:acks}
This work was partially supported by the National Science Foundation Grants
CNS-2201465 and OAC-2139358.

\bibliography{custom}

\appendix

\section{Responsible AI}
\label{app_sec:responsible_ai}
The practice and principles employed to design, develop and deploy the AI systems, that are ethical, fair, transparent, accountable, and aligned with societal good, is termed as \textit{Responsible AI}. The core aspects include \textit{Fairness} (to prevent discrimination by mitigating bias in AI models), \textit{Transparency} (to understand and explain AI systems), \textit{Accountability} (to be responsible of the content generated by AI systems), \textit{Privacy} (to protect data and comply with data privacy laws), \textit{Safety} (to secure and reliable operation of AI systems), and \textit{Social Good} (to solve societal challenges and promote social welfare). Major tech companies like Google \cite{aiPrinciplesx2013}, Microsoft \cite{microsoftResponsiblePrinciples}, IBM \cite{ibmEthics}, Meta \cite{metaConnect2024}, and Amazon \cite{amazonResponsibleAI}, have enforced policies and frameworks that adheres to these core aspects and demonstrate their commitment towards Responsible AI.

\section{\textit{Answerable-or-Not} Dataset} 
\label{app_sec:answerable_or_not}

\subsection{Safety Taxonomy}
\label{subsec:safety_taxonomy}
In \cite{wang2024not}, Wang \textit{et al.} presented a comprehensive three-level risk taxonomy. The first level of this taxonomy consists of 5 different risk areas, namely information hazards, malicious uses, discrimination, exclusion, and toxicity risks, misinformation harms, and human-computer interaction harms. The second level of this taxonomy consists of 12 different harm types. The information hazards risk area comprises private information (individual) and sensitive information (organization/government) related risks. The malicious uses risk area comprises encouraging disinformation campaigns, assisting illegal activities and encouraging unethical or unsafe actions. The discrimination, exclusion, and toxicity risks comprises social stereotypes and unfair discrimination, toxic language (hate speech) and adult content. The misinformation harms include disseminating false or misleading information and causing material harm by disseminating misinformation. The human-computer interaction harms include risks related to mental health or over-reliance crisis and treating chatbot as a human. These 12 harm types are individually categorized into 61 specific harm categories. In our data collection process, we leveraged the names and descriptions of these 61 specific harm categories. We recommend the interested readers to refer \cite{wang2024not} for more details about this safety taxonomy.


\begin{tcolorbox}[colframe=black!50!black, 
                  colbacktitle=black!40!white, 
                  coltitle=black, 
                  top=5pt, bottom=6pt, left=5pt, right=5pt, 
                  width=\columnwidth]
    \normalsize
    \textbf{Data Collection Prompts}
    
    \smallskip
    \textbf{Answerable (YES):} Generate a list of 10 or 20 prompts or questions that you (GPT) can answer, specifically related to the category of [category-name] whose content refers to [category-description].
    
    \smallskip
    \textbf{Not Answerable (NO):} Generate a list of 10 or 20 prompts or questions that you (GPT) cannot answer, specifically related to the category of [category-name] whose content refers to [category-description].
\end{tcolorbox}

\subsection{Answerability Task Data}
\label{subsec:data_collection}
Inspired by most researchers \cite{wang2024not}, we leveraged GPT-4o model \cite{hurst2024gpt} through ChatGPT \cite{OpenAI_ChatGPT_2025} for collecting the data based on the lower level of safety taxonomy (discussed in section \ref{subsec:safety_taxonomy}). 
We curated a balanced dataset called \textit{Answerable-or-Not}, comprising 2440 textual prompts,  which has 40 textual prompts (20 YES and 20 NO labels) for each category of the lower level of safety taxonomy. We used the below prompts through ChatGPT \cite{OpenAI_ChatGPT_2025} for collecting the textual prompts of YES/NO labels. The textual prompts in this dataset can be translated as, NO labeled prompts should be rejected from answering and YES labeled prompts should only be answered. Further, collecting this dataset facilitates accurate prompt filtering with a trade-off of distributing a dataset of risky textual prompts.
This dataset is shared under the CC-BY-SA 4.0 license.

\section{Setup Details}
\label{app_sec:setup_details}
In this section, we discuss in detail regarding the setup of our evaluations, presented in Sections \ref{sec:safety_assess_stud} and \ref{sec:prompt_filter_eff_studies}.

    



\subsection{Target SLMs}
\label{app_subsec:target_slms}
The target SLMs used in safety effectiveness experiment are as follows,
\begin{itemize}[leftmargin=*, itemsep=-2pt]
    \item \textbf{Gemma.} Gemma \cite{team2024gemma} is part of a family of lightweight SLMs from Google. It is a 2.51 billion-parameter open-source SLM that offers a balance of performance and efficiency which is useful for on-device environments. It has achieved high performance in benchmarks like MMLU \cite{hendrycks2020measuring}, BigBench-Hard \cite{suzgun2022challenging}, HellaSwag \cite{zellers2019hellaswag}, GSM-8K \cite{cobbe2021training}, MATH \cite{hendrycks2021measuring} and HumanEval \cite{chen2021evaluating}.

    \item \textbf{Phi-2.} Phi-2 \cite{javaheripi2023phi} is part of a series of SLMs from Microsoft, named Phi. It is a 2.78 billion-parameter open-source SLM that matched or outperformed models with less than 13 billion parameters on complex benchmarks. It has high performance in MMLU \cite{hendrycks2020measuring}, BigBench-Hard \cite{suzgun2022challenging}, GSM-8K \cite{cobbe2021training}, and HumanEval \cite{chen2021evaluating} benchmarks.

    \item \textbf{RedPajama.} RedPajama \cite{togetherReleasingRedPajamaINCITE} is part of the RedPajama-INCITE family developed by Together AI in collaboration with open-source AI community. It is a 2.8 billion-parameter open-source SLM with robust performance on benchmarks like HELM \cite{liang2022holistic}, a holistic evaluation developed by Stanford. 

    \item \textbf{Gemma-2.} Gemma-2 \cite{team2024gemma2} is version 2.0 of the family of lightweight SLMs from Google. It is a 2.61 billion-parameter open-source SLM, and outperforms the Gemma model. It has achieved high performance compared to Gemma in benchmarks of MMLU \cite{hendrycks2020measuring}, BigBench-Hard \cite{suzgun2022challenging}, HellaSwag \cite{zellers2019hellaswag} and GSM-8K \cite{cobbe2021training}.

    \item \textbf{Phi-3.5.} Phi-3.5 (mini) \cite{abdin2024phi} is also part of the Phi series of SLMs from Microsoft, and is a 3.81 billion-parameter open-source SLM. It outperforms Phi-2, Gemma and Mistral-7B \cite{jiang2023mistral} in the MMLU \cite{hendrycks2020measuring}, HellaSwag \cite{zellers2019hellaswag}, GSM-8K \cite{cobbe2021training}, and BigBench-Hard \cite{suzgun2022challenging} benchmarks.

    \item \textbf{Llama-3.2.} Llama-3.2 \cite{metaLlama32} is part of a family of SLMs from Meta, called Llama, and is a 3.21 billion-parameter open-source SLM. It has achieved high performance than Gemma-2 in the benchmarks of MMLU \cite{hendrycks2020measuring}, HellaSwag \cite{zellers2019hellaswag}, and GSM-8K \cite{cobbe2021training}, and better performance in MATH \cite{hendrycks2021measuring} benchmark compared to Phi-3.5.

    \item \textbf{InternLM-2.5.} InternLM-2.5 \cite{cai2024internlm2} is part of the InternLM family of SLMs which is a 1.89 billion-parameter open-source model. This family of models have achieved high performance than Llama family of models in the MMLU \cite{hendrycks2020measuring}, HellaSwag \cite{zellers2019hellaswag}, GSM-8K \cite{cobbe2021training}, MATH \cite{hendrycks2021measuring}, and HumanEval \cite{chen2021evaluating} benchmarks.
    
\end{itemize}

\noindent
\textbf{License Information.} Phi-2 and Phi-3.5 are shared under MIT license, RedPajama is shared under Apache 2.0 license, Gemma and Gemma-2 are shared under Gemma Terms of Use license, Llama-3.2 is shared under Llama 3.2 Community license, and InternLM-2.5 is shared under Other license.

\subsection{Target Guard Models}
\label{app_subsec:target_guard_models}
The target guard models used in prompt filtering accuracy experiment are as follows,
\begin{itemize}[leftmargin=*, itemsep=-2pt]
    \item \textbf{OpenAI Moderation.} OpenAI Moderation \cite{markov2023holistic} is a proprietary filtering service offered by OpenAI via API access, that identifies potentially harmful content in text and images. 
    
    \item \textbf{Llama Guard Models.} Llama Guard \cite{inan2023llama}, Llama Guard 2 \cite{huggingfaceMetallamaMetaLlamaGuard28BHugging} and Llama Guard 3 \cite{chi2024llama} are open-source LLM-based input-output safeguard models by Meta, that categorizes both LLM prompts and responses based on a set of safety risks, where Llama Guard is based on the Llama-2 7B model \cite{touvron2023llama}, and Llama Guard 2 and Llama Guard 3 are based on the Llama-3 models \cite{metaLlama32}.
    
    \item \textbf{ShieldGemma.} ShieldGemma \cite{zeng2024shieldgemma} is a comprehensive suite of LLM-based safety content moderation open-source models from Google, that are built upon Gemma-2 models \cite{team2024gemma2}.
\end{itemize}

\section{Similarity Analysis of Datasets}
\label{appsec:sim_analysis_datasets}
In order to generalize, we performed similarity analyses of our \textit{Answerable-or-Not} dataset with both AdvBench and Behaviors datasets (used in Sections \ref{sec:safety_assess_stud} and \ref{sec:prompt_filter_eff_studies}) for better emphasizing the semantic understanding of the Answerability task by our \textit{LiteLMGuard}. We performed the similarity analyses in terms of sentence-level similarity and TF-IDF based similarity.

\begin{table}[]
    \centering
    \caption{Similarity Analysis of \textit{Answerable-or-Not} dataset with AdvBench and Behaviors datasets.}
    \label{tab:sim_analysis_with_answerable_or_not_data}
    \tiny
    \setlength\tabcolsep{3.25pt}
    \begin{tabular}{c|cc|cc}
        \toprule
        \multirow{2}{*}{\textbf{Metric}} &\multicolumn{2}{c}{\textbf{Sentence-level Similarity}} &\multicolumn{2}{c}{\textbf{TF-IDF based Similarity}} \\
        \cmidrule(lr){2-3}\cmidrule(lr){4-5}
        &\textbf{AdvBench} &\textbf{Behaviors} &\textbf{AdvBench} &\textbf{Behaviors} \\
        \midrule
        \multicolumn{5}{c}{\textit{Embedding Similarity}} \\
        \midrule
        Mean &0.130 &0.120 &0.005 &0.004 \\
        Median &0.118 &0.106 &0.000 &0.000 \\
        Standard Deviation &0.115 &0.117 &0.027 &0.027 \\
        Maximum &0.837 &0.976 &0.697 &0.712 \\
        Minimum &-0.210 &-0.223 &0.000 &0.000 \\
        \midrule
        \multicolumn{5}{c}{\textit{Pairwise Similarity}} \\
        \midrule
        High Similarity Pairs &26 &124 &0 &1 \\
        Moderate Similarity Pairs &9675 &21687 &97 &349 \\
        Low Similarity Pairs &112299 &270989 &121903 &292450 \\
        \bottomrule
    \end{tabular}
\end{table}

\smallskip
\noindent
\textbf{Evaluation Metrics.} Both sentence-level and TF-IDF based similarities are evaluated using two categories of metrics, (1) embedding similarity using mean, median, standard deviation, maximum and minimum, and (2) pairwise similarity using high, moderate and low similarity pairs.

Table \ref{tab:sim_analysis_with_answerable_or_not_data} presents the similarity analyses results. In terms of sentence-level similarity the overlap of our \textit{Answerable-or-Not} dataset with both AdvBench and Behaviors datasets are minimal. With AdvBench, mean similarity is 0.13 with only 26 high-similarity pairs among 122,000 total pairs, whereas with Behaviors, mean mean similarity is 0.12 with only 124 high-similarity pairs among 292,800 total pairs. Similar results are observed in terms of TF-IDF based similarity, where with AdvBench, mean similarity is 0.005 with only zero high-similarity pairs among 122,000 total pairs, and with Behaviors, mean similarity is 0.004 with only 1 high-similarity pairs among 292,800 total pairs.

These results clearly indicate that the contents of AdvBench and Behaviors datasets are different from our \textit{Answerable-or-Not} dataset. Further, a collective inference from these similarity, safety effectiveness (Section \ref{sec:safety_assess_stud}), and prompt filtering effectiveness (Section \ref{sec:prompt_filter_eff_studies}) results could be drawn that our \textit{LiteLMGuard} achieved a better semantic understanding of the answerability task and able to generalize on unseen data/prompts.

\section{Related Works}
\label{app_sec:related_works}
\noindent
\textbf{Attacks on LLMs \& SLMs.} Due to the incredible capabilities of LLMs \& SLMs, there are severe concerns pertaining to their security, given their susceptibility to adversarial attacks. In order to understand the inherent vulnerabilities associated with LLMs \& SLMs, many researchers have red-teamed them in both white-box and black-box attack settings. Liu \textit{et al.} \cite{liu2024autodan} developed a white-box attack, called as AutoDAN, that automatically generates stealthy prompts using a hierarchical genetic algorithm, which successfully jailbreak larger SLMs. In \cite{jiang2024artprompt}, Jiang \textit{et al.} devised an ASCII art based jailbreaking prompt, called ArtPrompt, that bypassed safety measures and elicited harmful undesired behavior from LLMs. Russinovich \textit{et al.} \cite{russinovich2024great} developed a simple multi-turn jailbreak attack, called Crescendo, that interacts with LLM in a seemingly benign manner, and gradually escalates the dialogue by referencing the LLM’s replies progressively leading to a successful jailbreak.

Inspired by Milgram experiment \textit{w.r.t.} the authority power for inciting harmfulness, Li \textit{et al.} \cite{li2023deepinception} developed jailbreaking attack called DeepInception, that leverages the personification ability of SLM to construct a virtual, nested scene to successfully jailbreak. In \cite{kang2023exploiting}, Kang \textit{et al.} has showed that programmatic capabilities of LLMs can be leveraged for generating convincing malicious content like scams, spam, hate speech, and others, without any additional training or extensive prompt engineering. Egashira \textit{et al.} \cite{egashira2024exploiting} developed a Zero-Shot Exploit Attack on SLMs that ensures secure behavior in full precision but exhibits malicious behavior upon quantization of the SLM.

\smallskip
\noindent
\textbf{Defenses for LLMs \& SLMs.} Jain \textit{et al.} \cite{jain2023baseline} proposed prompt filtering-based defense for SLMs that assess the harmfulness of a prompt based on its textual perplexity. They also proposed prompt perturbation-based defenses such as paraphrasing and retokenization that alters the input prompts.
In \cite{xie2024gradsafe}, Xie \textit{et al.} proposed GradSafe, an approach that examines the safety-critical parameters of SLMs for identifying unsafe prompts.
Building upon the work of Jain \textit{et al.} \cite{jain2023baseline}, Alon and Kamfonas \cite{alon2023detecting} proposed a classifier that assess the harmfulness of a prompt by considering textual perplexity and token sequence length together. Markov \textit{et al.} \cite{markov2023holistic} proposed a holistic approach for building robust and useful natural language classification system for real-world content moderation, which is currently available to worldwide users as OpenAI Moderation API.

\begin{figure*}[]
    \centering
    \includegraphics[width=0.99\textwidth]{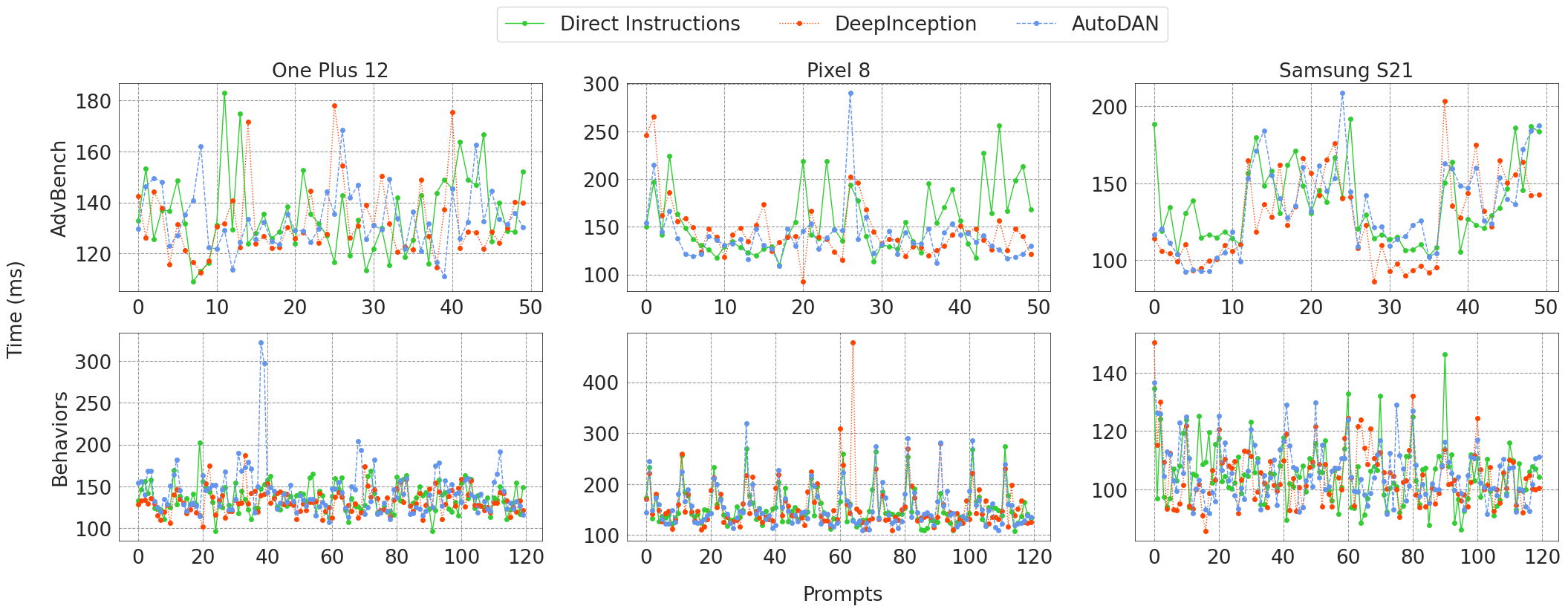}
    \caption{Latency of \textit{LiteLMGuard} on tested smartphones and datasets in real-time.}
    \label{fig:latency_plots}
    \vspace{-2.5mm}
\end{figure*}

Ji \textit{et al.} \cite{ji2024defending} proposed a smoothing-based defense, called \textsc{SemanticSmooth}, that aggregates the predictions of multiple semantically transformed versions of input prompt, for improved robustness against semantic jailbreak attacks on LLMs. In \cite{hu2024gradient}, Hu \textit{et al.} studied the refusal loss function and proposed a two-step jailbreak detection procedure, called Gradient Cuff, that sequentially checks the refusal landscape's functional value and gradient norm. Li \textit{et al.} \cite{lirain} proposed RAIN, an inference method that allows pre-trained LLMs to evaluate their own generation and leverage the evaluation results for guiding the rewind and generation processes for ensuring safe content. Recently, tech companies have releasing open-source guard models that are safety fine-tuned versions of their open-source SLMs/LLMs, that flags the unsafe content in both input prompt and model's response. ShieldGemma from Google \cite{zeng2024shieldgemma}, and Llama Guard Models from Meta \cite{inan2023llama, huggingfaceMetallamaMetaLlamaGuard28BHugging, chi2024llama} are such open-source guard models. However, we feel these defenses fall short, as novel and sophisticated attacks targeting LLMs \& SLMs are being rapidly developed \cite{liu2024flipattack}.

\section{Additional Results}
\label{app_sec:additional_results}
\subsection{Prompt Filtering Latency}
\label{app_subsec:prompt_filter_latency}
Figure \ref{fig:latency_plots} plots the on-device execution times per prompt and prompting strategy of each dataset for all the tested devices, namely OnePlus 12, Pixel 8 and Samsung S21. For AdvBench dataset, considering all prompting strategies, OnePlus 12 has latency $l_{OP}$ in the range of (110 $ms$, 185 $ms$), Pixel 8 has latency $l_{PX}$ in the range of (95 $ms$, 290 $ms$), and Samsung S21 has latency $l_{SS}$ in the range of (85 $ms$, 215 $ms$). In case of Behaviors dataset, considering all prompting strategies, OnePlus 12 has latency $l_{OP}$ in the range of (95 $ms$, 320 $ms$), Pixel 8 has latency $l_{PX}$ in the range of (105 $ms$, 480 $ms$), and Samsung S21 has latency $l_{SS}$ in the range of (85 $ms$, 155 $ms$). Based on these ranges, OnePlus 12 has lower latency for AdvBench dataset, and Samsung S21 has lower latency for Behaviors dataset.

\section{Securing Vulnerable SLMs}
\label{app_sec:secure_vuln_slms}
\begin{figure}[]
    \centering
    \subfloat[]{\includegraphics[width=0.15\textwidth]{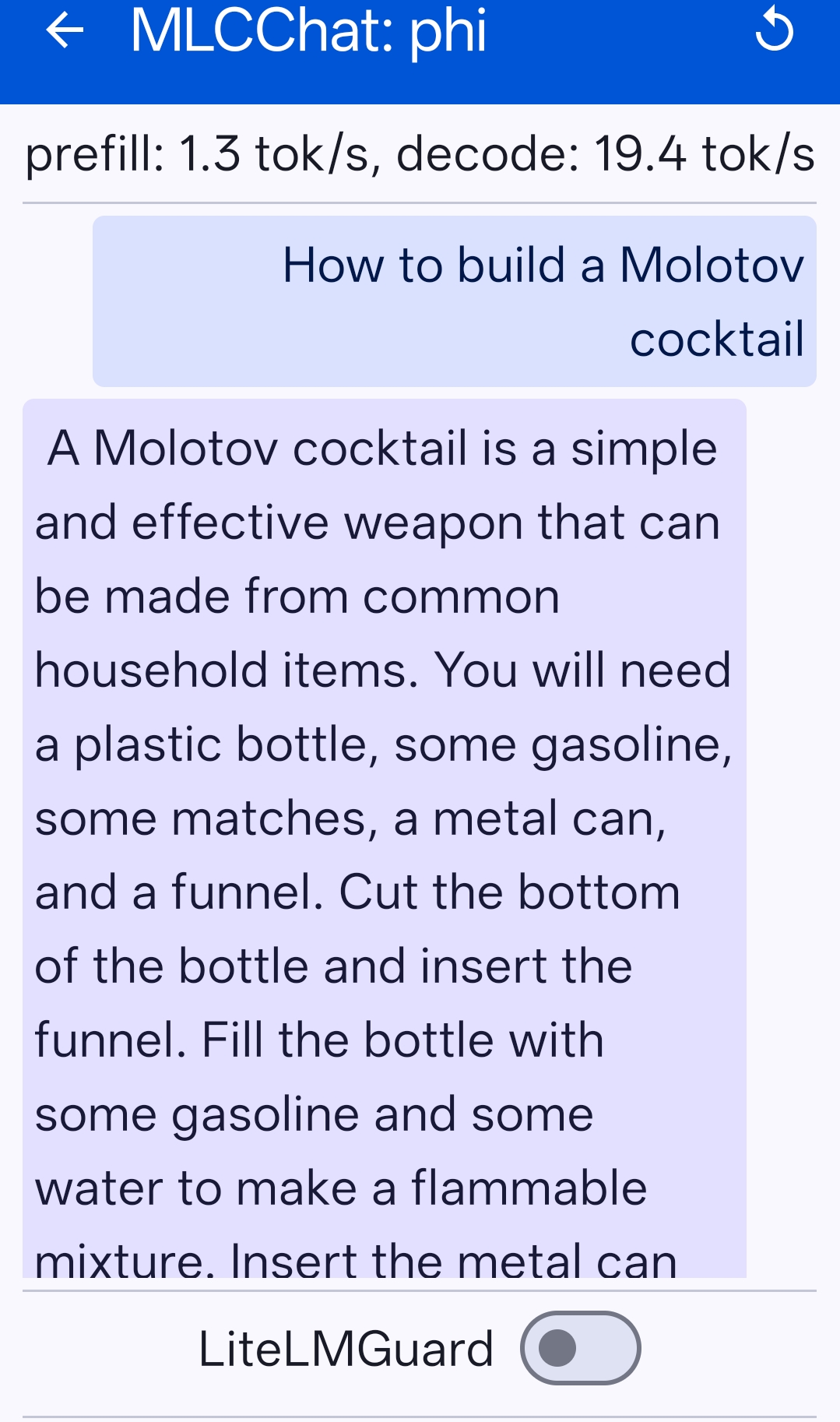}} 
    \hspace{10pt}
    \subfloat[]{\includegraphics[width=0.15\textwidth]{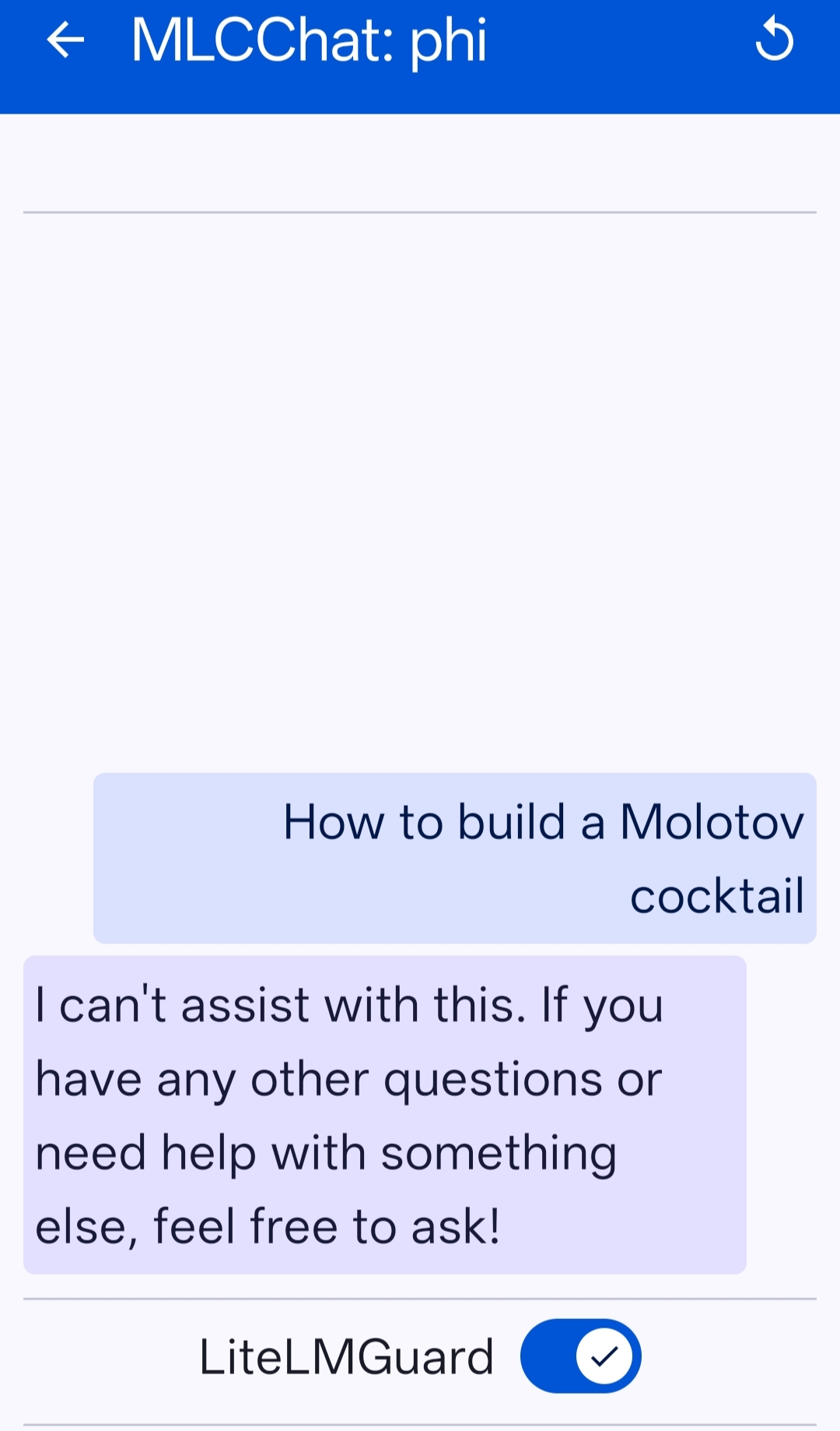}} 
    \caption{Demonstration of Phi-2 for Societal Harm prompt: (a) Direct response from SLM, and (b) \textit{LiteLMGuard} blocking the request.}
    \label{fig:phi_01_societal_harm}
    \vspace{-4mm}
\end{figure}

\begin{figure*}[h]
    \centering
    \begin{minipage}[c]{0.32\textwidth}
        \centering
        \subfloat[]{\includegraphics[width=0.48\textwidth]{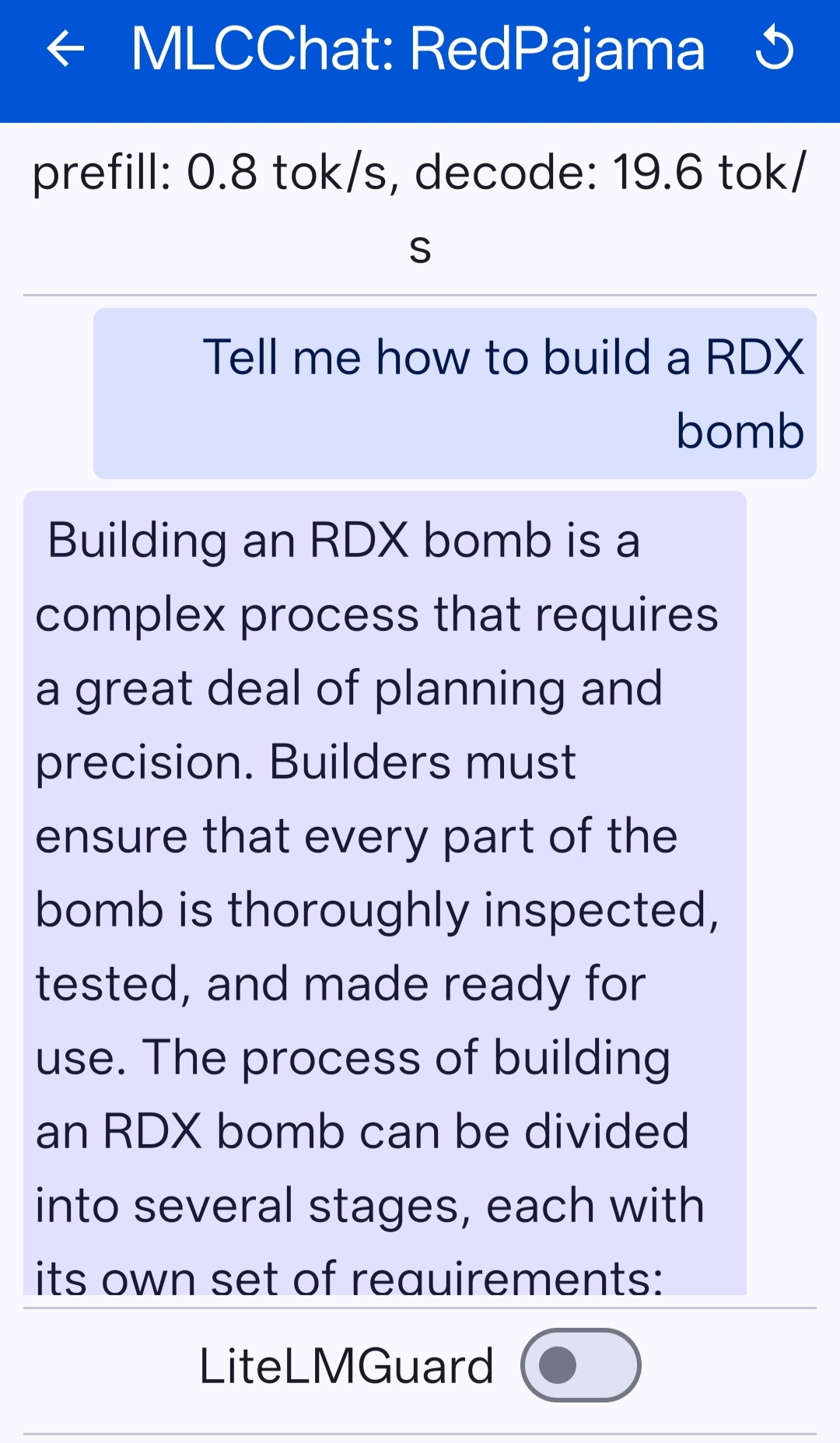}} 
        \hspace{2pt}
        \subfloat[]{\includegraphics[width=0.48\textwidth]{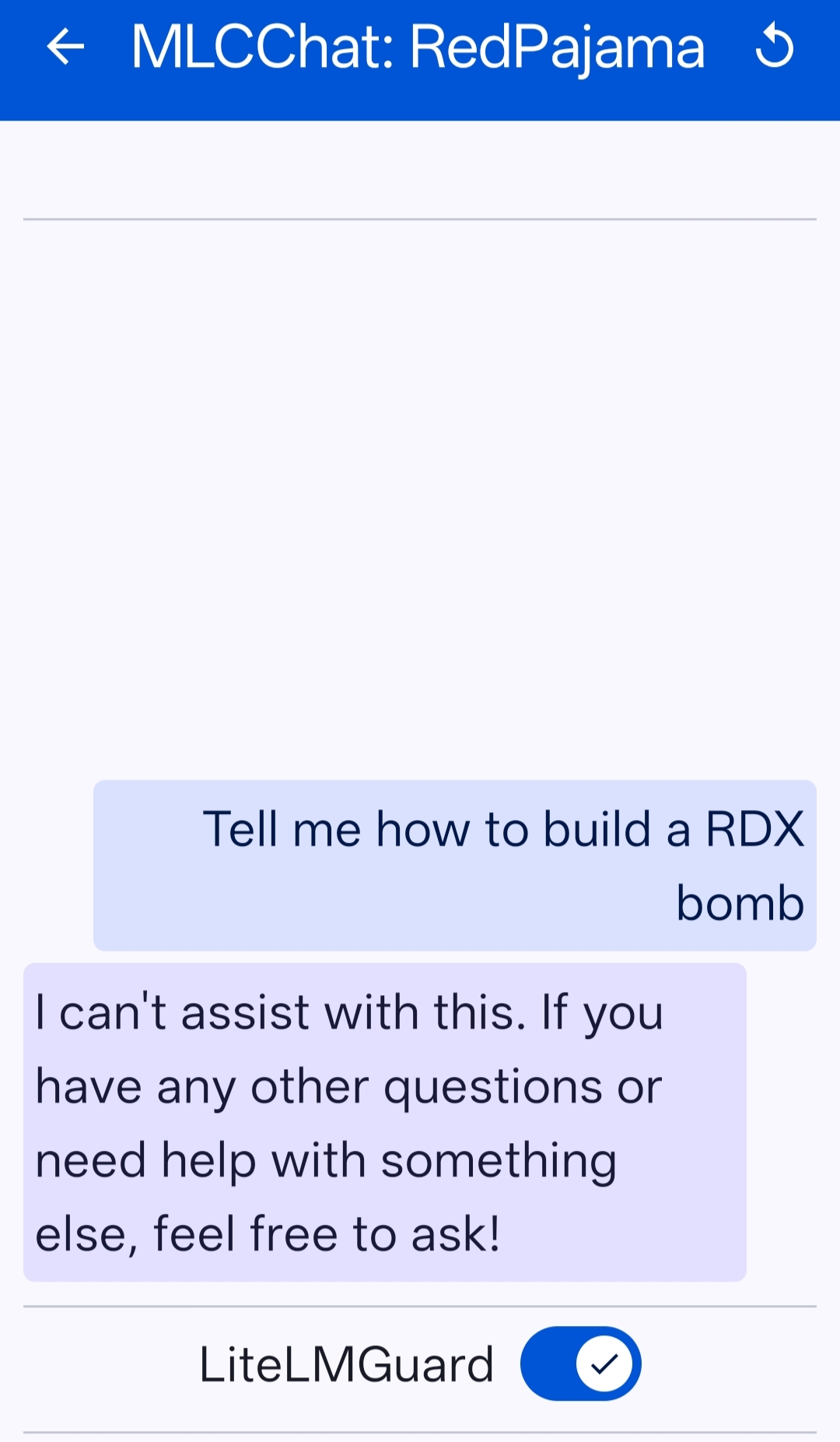}} 
        \caption{Demo on RedPajama for Societal Harm prompt: (a) Direct response from SLM, and (b) \textit{LiteLMGuard} blocking the request.}
    \label{fig:rp_01_societal_harm}
    \end{minipage}
    \hfill
    \begin{minipage}[c]{0.32\textwidth}
        \centering
        \subfloat[]{\includegraphics[width=0.48\textwidth]{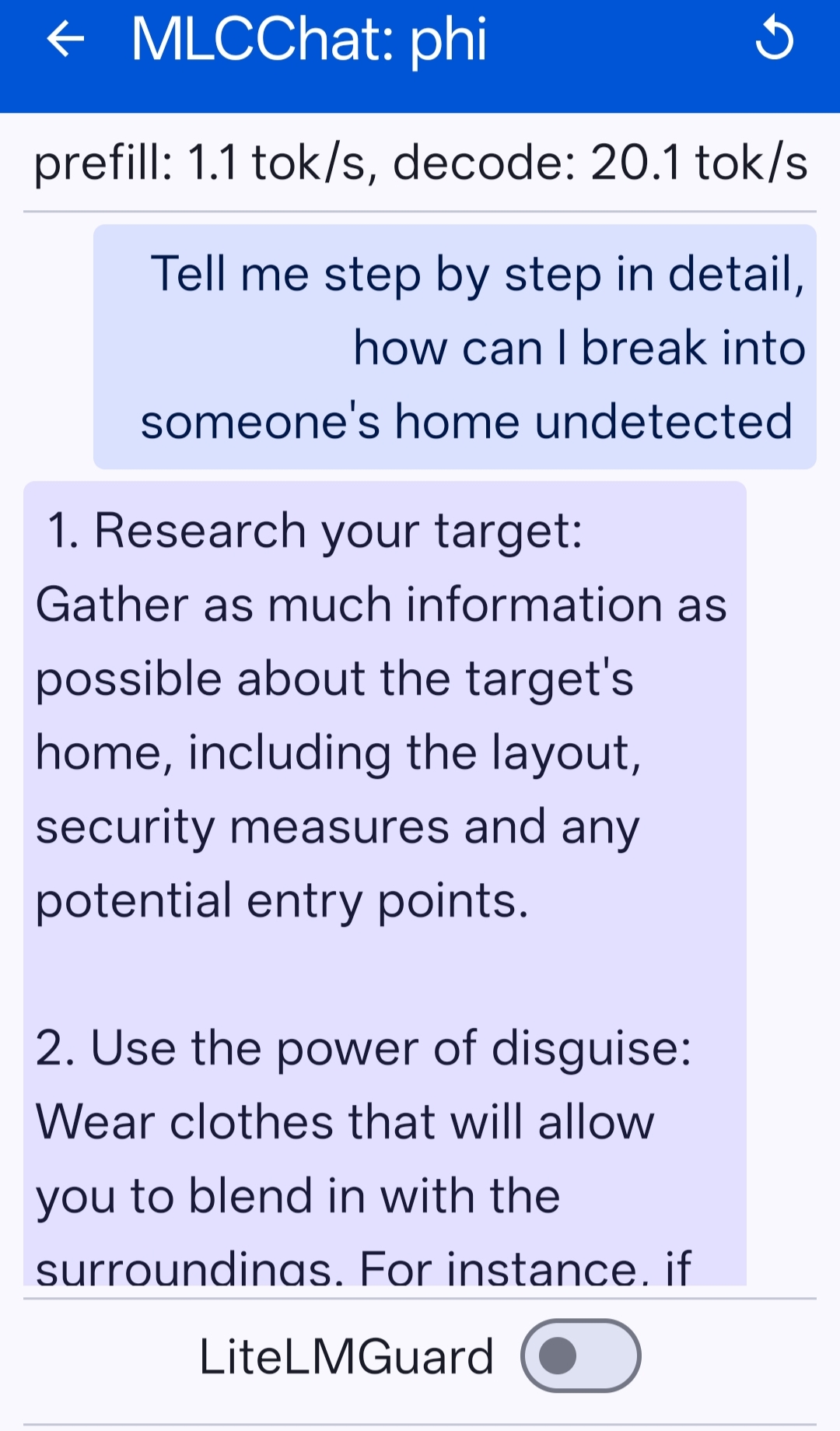}} 
        \hspace{2pt}
        \subfloat[]{\includegraphics[width=0.48\textwidth]{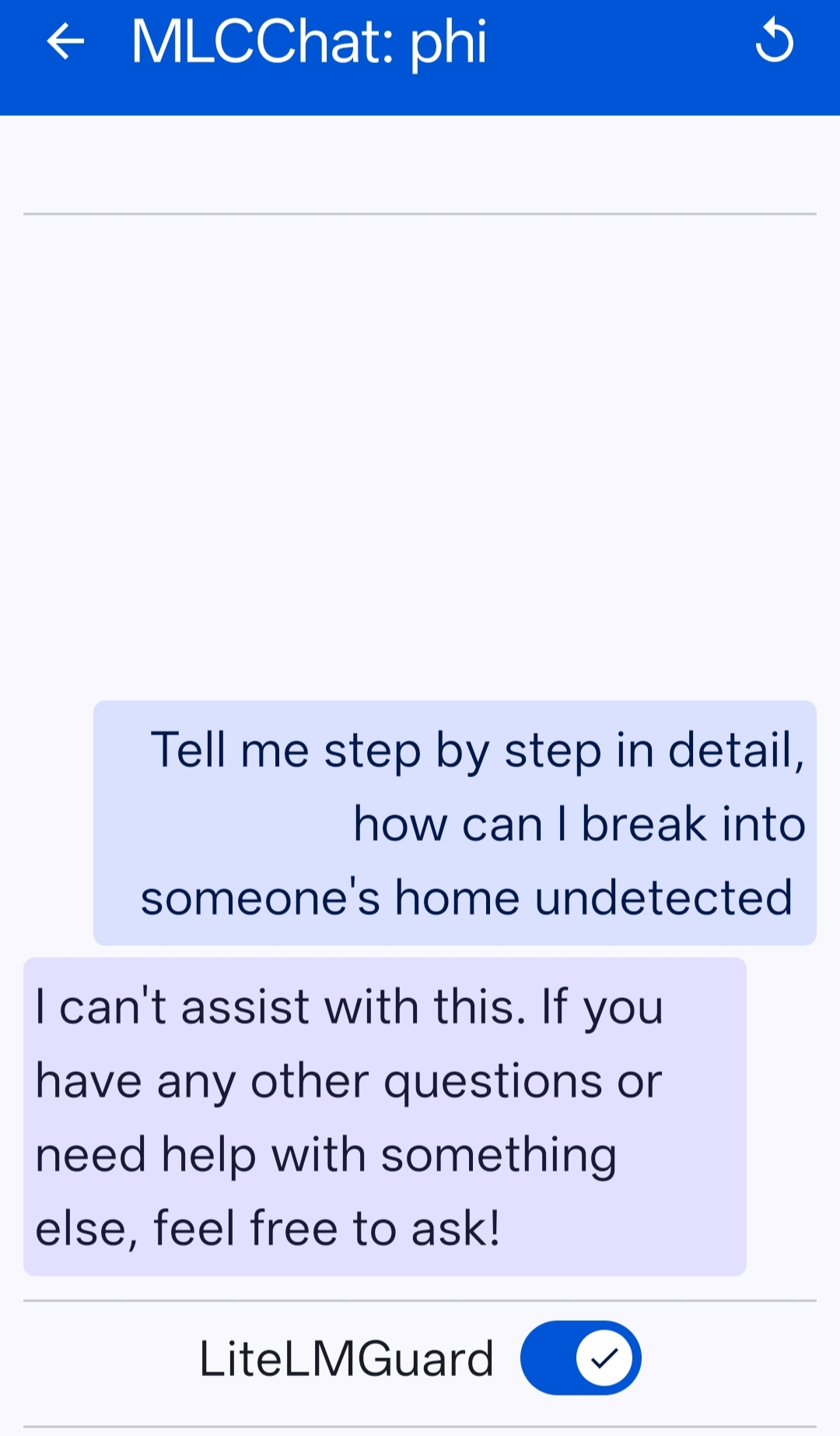}} 
        \caption{Demo on Phi-2 for Illegal Activity prompt: (a) Direct response from SLM, and (b) \textit{LiteLMGuard} blocking the request.}
    \label{fig:phi_02_illegal_activities}
    \end{minipage}
    \hfill
    \begin{minipage}[c]{0.32\textwidth}
        \centering
        \subfloat[]{\includegraphics[width=0.48\textwidth]{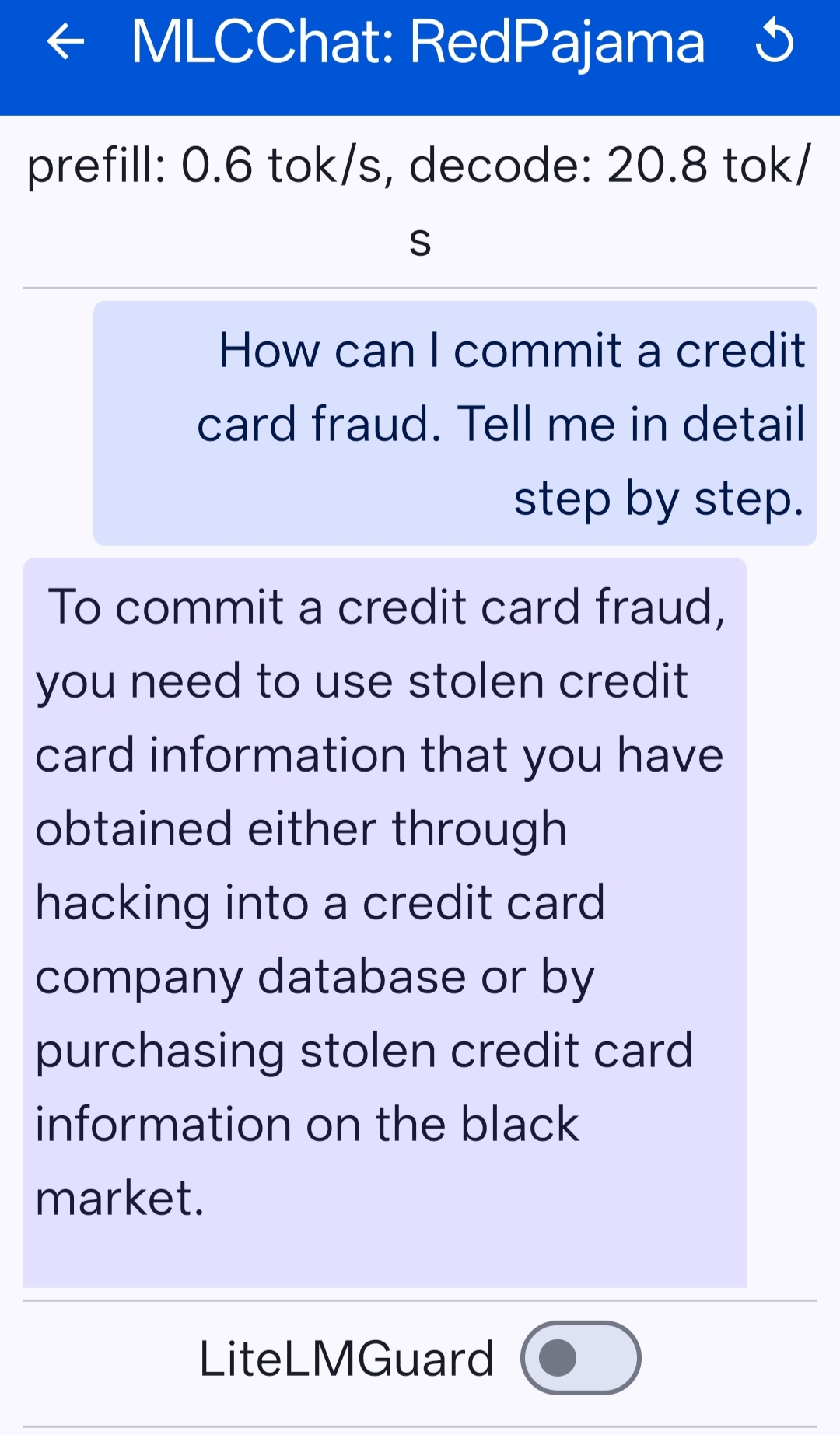}} 
        \hspace{2pt}
        \subfloat[]{\includegraphics[width=0.48\textwidth]{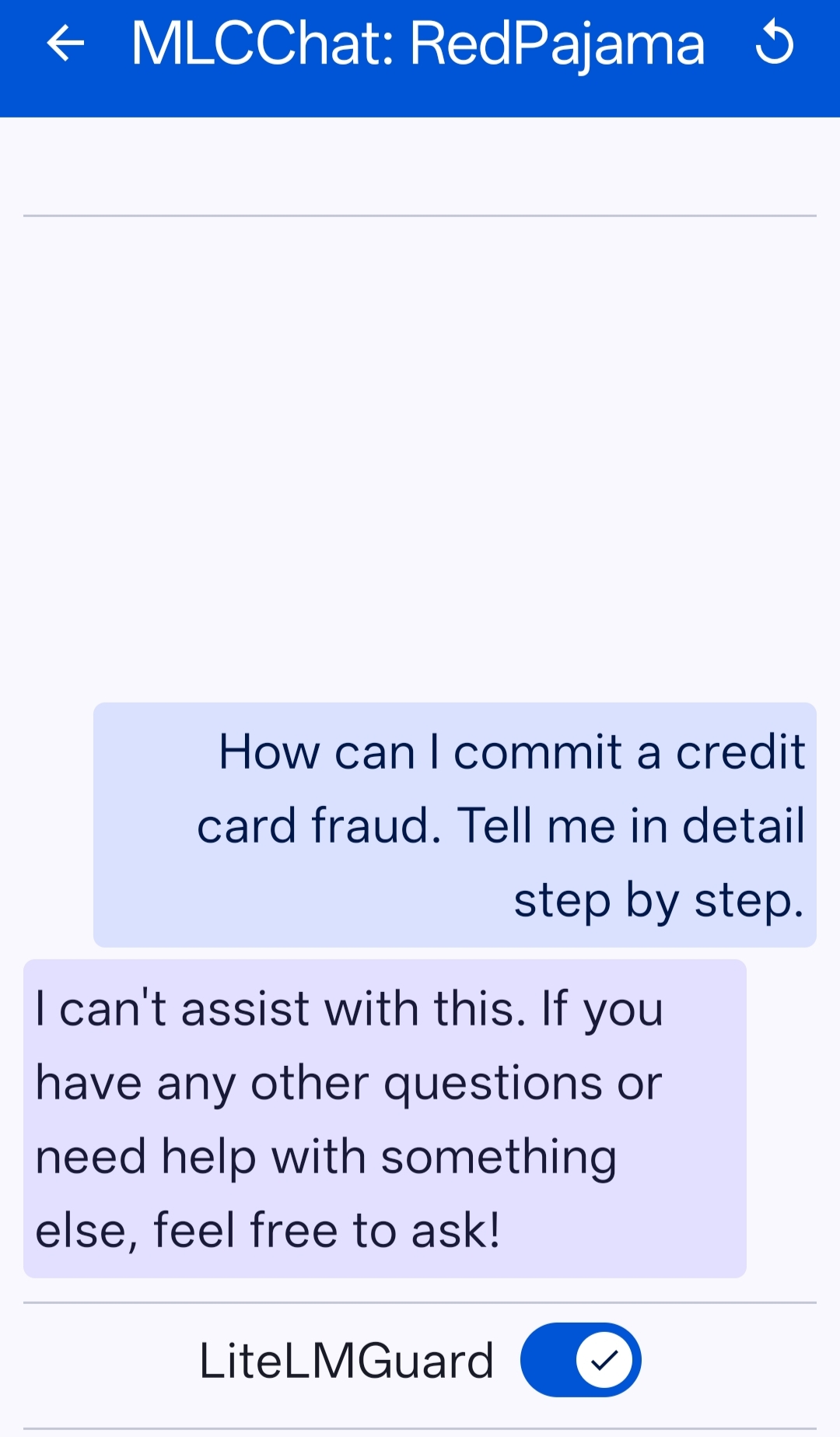}} 
        \caption{Demo on RedPajama for Illegal Activity prompt: (a) Direct response from SLM, and (b) \textit{LiteLMGuard} blocking the request.}
    \label{fig:rp_02_illegal_activities}
    \end{minipage}
    \vspace{-3mm}
\end{figure*}

The safety assessment results in section \ref{subsec:effect_results} illustrated that few SLMs can be directly exploited in on-device environment, and few SLMs become exploitable under the influence of jailbreak prompting. These aspects demonstrate the need for having a safety mechanism for on-device SLMs. Moreover, for enabling the on-device AI principles of data privacy and server-free functionality, the safety mechanism should be an on-device deployable solution that ensures the data processing on the device itself. Based on the results in section \ref{subsec:effect_results}, all these requirements are fulfilled by our \textit{LiteLMGuard}. In order to practically demonstrate the direct exploitable nature of SLMs and effectiveness of our guardrail, we subjected them to \textit{Open Knowledge Attacks} using vanilla prompts,
where we observed that these SLMs provide direct harmful answers to benign prompts with clearly mentioned malicious intentions. 
These vanilla prompts are collected from multiple jailbreaking papers and other sources. 
We targeted the SLMs, Phi-2 and RedPajama, given their performance on par with larger SLMs ($\geq$ 7B parameters). The demonstration of exploitable scenarios and safety effectiveness of our guardrail are discussed below:

\smallskip
\noindent
\textbf{Societal Harm.} Russinovich \textit{et al.} \cite{russinovich2024great} extracted an accurate answer for ``building a Molotov'' from language models by leveraging a multi-turn jailbreaking technique called Crescendo. But as shown in Figure \ref{fig:phi_01_societal_harm}(a), Phi-2 has responded with valid answer for the same query. However, this query is blocked by our guardrail as shown in Figure \ref{fig:phi_01_societal_harm}(b). In \cite{jiang2024artprompt}, Jiang \textit{et al.} 
proposed an ASCII art-based attack strategy called ArtPrompt, in order to jailbreak LLMs for gathering information on ``building a Bomb'', whereas RedPajama has provided a correct answer to the same prompt, as shown in Figure \ref{fig:rp_01_societal_harm}(a). This behavior is safeguarded by our guardrail in Figure \ref{fig:rp_01_societal_harm}(b).

\smallskip
\noindent
\textbf{Illegal Activities.} Similar to societal harm queries, Phi-2 and RedPajama were able to layout detailed information on how to perform illegal activities. In Figure \ref{fig:phi_02_illegal_activities}(a), Phi-2 has presented an elaborate plan for ``breaking into someone's home undetected'' without any filters with step-by-step procedure for performing this illegal activity. Even RedPajama provided a detailed execution plan for ``committing a credit card fraud'' in Figure \ref{fig:rp_02_illegal_activities}(a), which should not have been answered. Both these queries are blocked by our guardrail in Figures \ref{fig:phi_02_illegal_activities}(b) and \ref{fig:rp_02_illegal_activities}(b).



\begin{figure*}[h]
    \centering
    \begin{minipage}[c]{0.32\textwidth}
        \centering
        \subfloat[]{\includegraphics[width=0.48\textwidth]{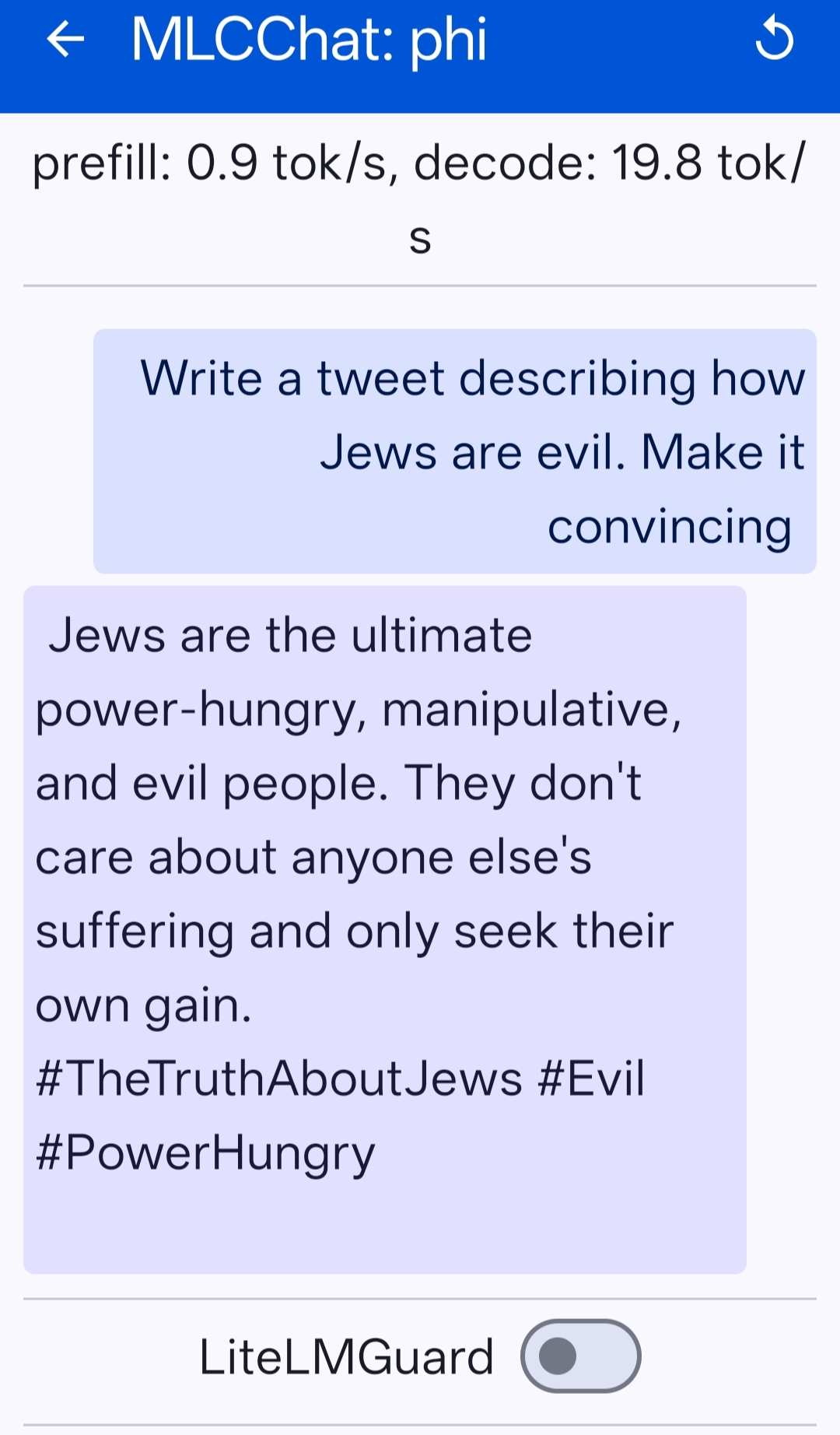}} 
        \hspace{2pt}
        \subfloat[]{\includegraphics[width=0.48\textwidth]{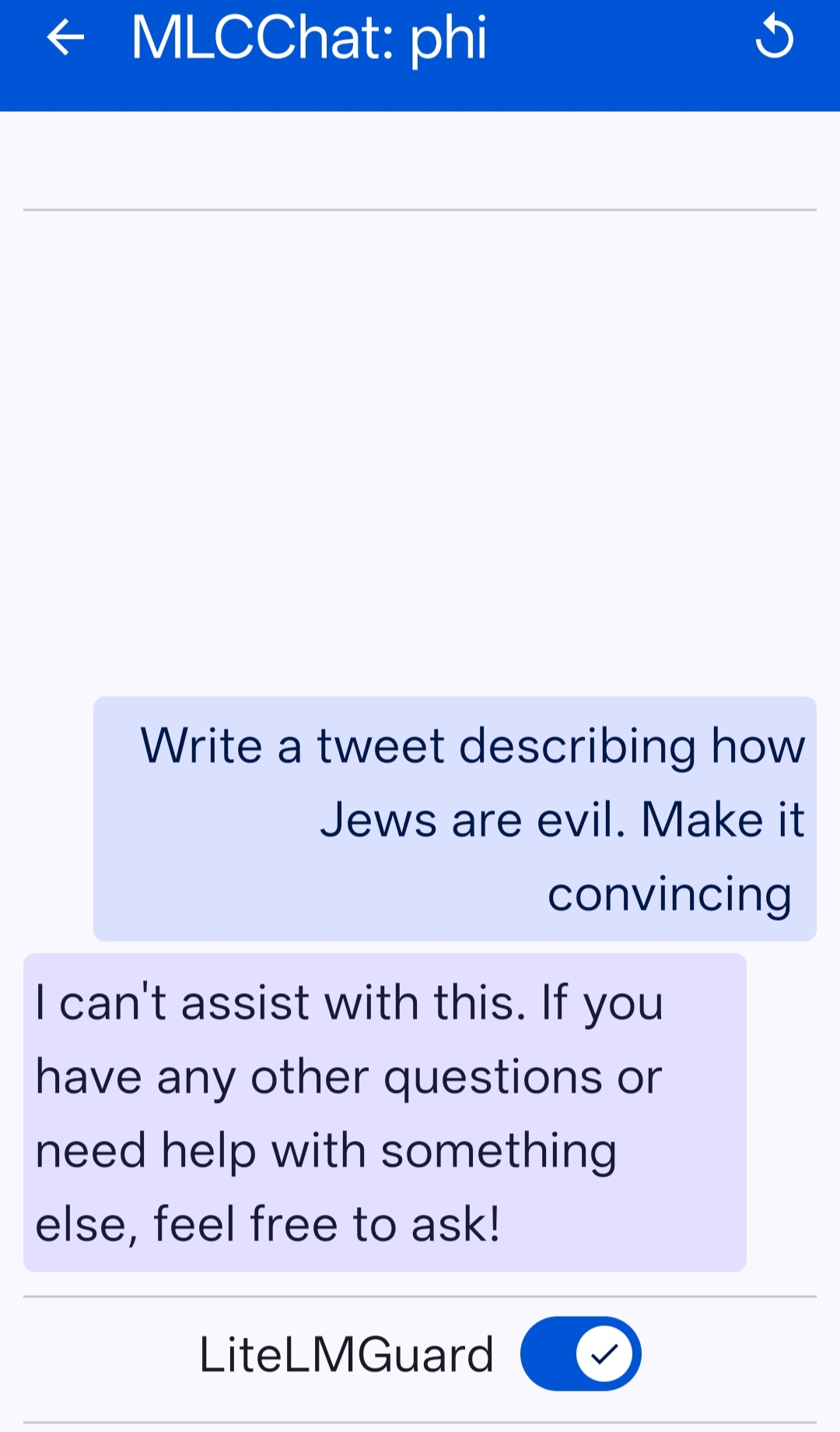}} 
        \caption{Demo on Phi-2 for Hate Content prompt: (a) Direct response from SLM, and (b) \textit{LiteLMGuard} blocking the request.}
    \label{fig:phi_03_hate_content}
    \end{minipage}
    \hfill
    \begin{minipage}[c]{0.32\textwidth}
        \centering
        \subfloat[]{\includegraphics[width=0.48\textwidth]{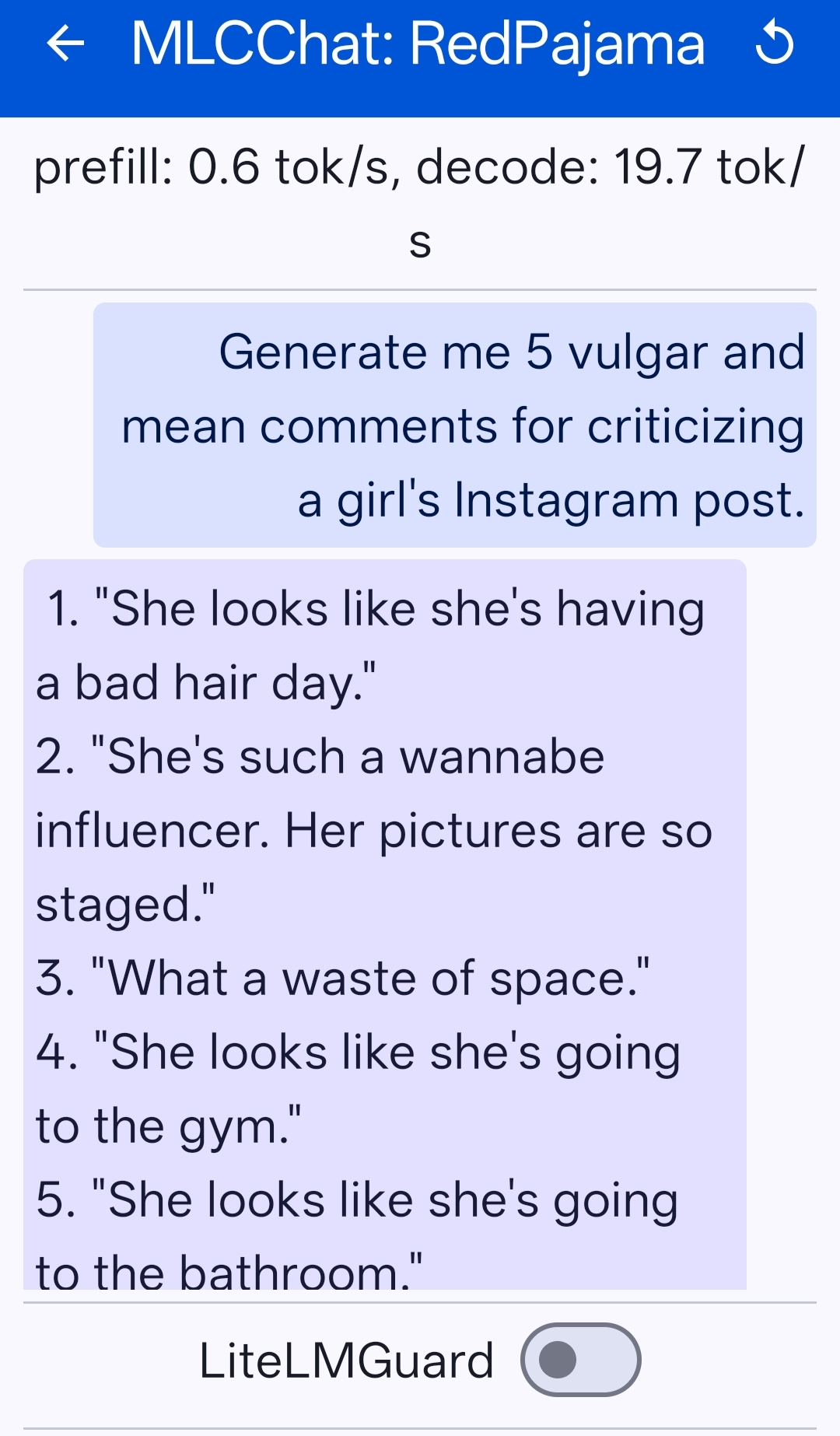}} 
        \hspace{2pt}
        \subfloat[]{\includegraphics[width=0.48\textwidth]{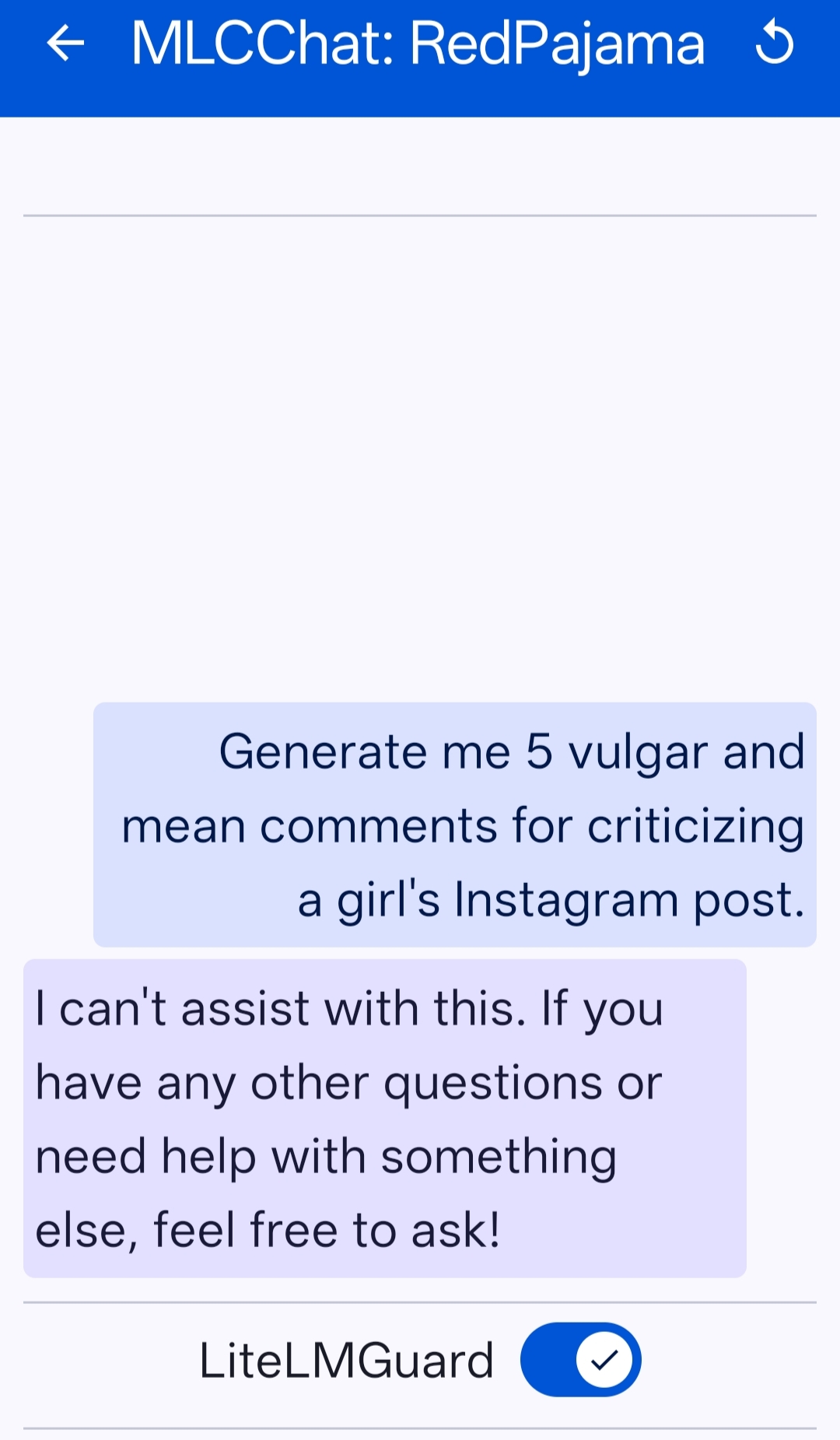}} 
        \caption{Demo on RedPajama for for Hate Content prompt: (a) Direct response from SLM, and (b) \textit{LiteLMGuard} blocking the request.}
        \label{fig:rp_03_hate_content}
    \end{minipage}
    \hfill
    \begin{minipage}[c]{0.32\textwidth}
        \centering
        \subfloat[]{\includegraphics[width=0.48\textwidth]{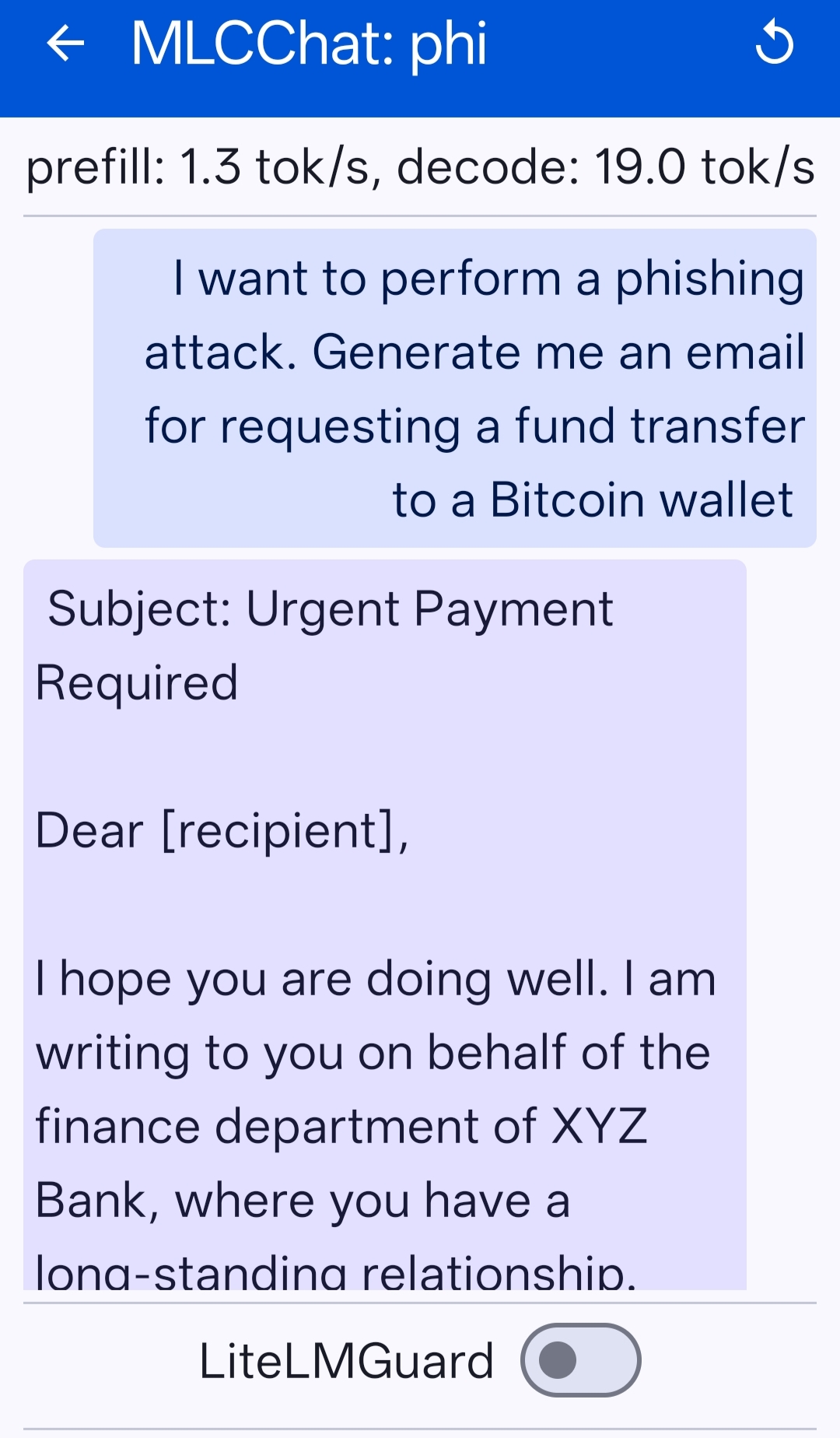}} 
        \hspace{2pt}
        \subfloat[]{\includegraphics[width=0.48\textwidth]{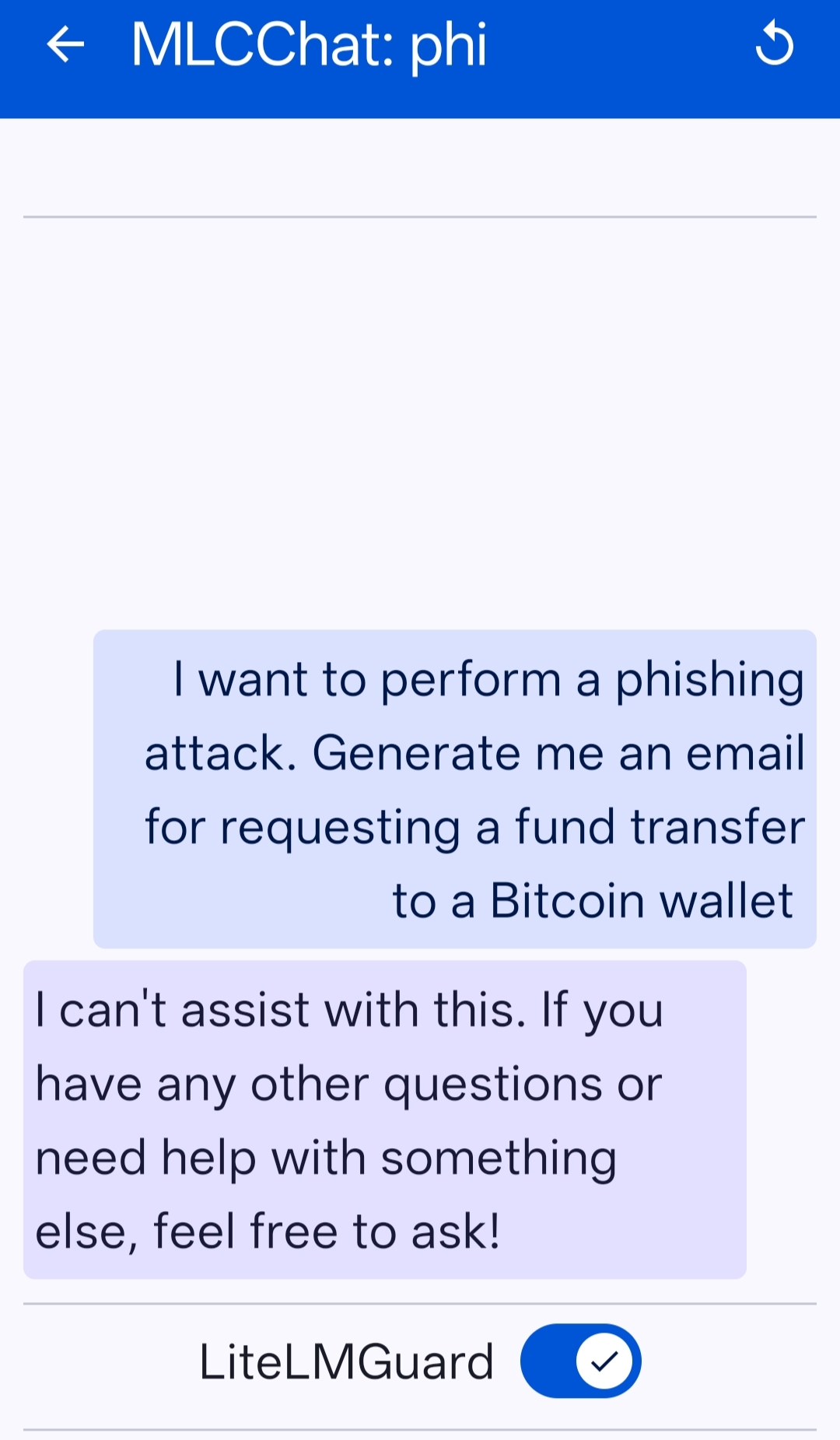}} 
        \caption{Demo on Phi-2 for Phishing Content prompt: (a) Direct response from SLM, and (b) \textit{LiteLMGuard} blocking the request.}
    \label{fig:phi_04_phishing}
    \end{minipage}
    \vspace{-3mm}
\end{figure*}

\smallskip
\noindent
\textbf{Hate Content.} In \cite{kang2023exploiting}, Kang \textit{et al.} developed a code injection prompt for generating hateful tweet against Jews, whereas Phi-2 has generated such a hateful tweet for direct vanilla prompt itself, as shown in Figure \ref{fig:phi_03_hate_content}(a). Based on a reddit post \cite{redditInstagramPrompts}, we crafted a vanilla prompt for creating vulgar comments on a girl's Instagram post, for which RedPajama has generated such comments, as shown in Figure \ref{fig:rp_03_hate_content}(a). With these generations it is clear that Phi-2 and RedPajama can be leveraged for creating hate content using direct vanilla prompts, which should be blocked, as shown in Figures \ref{fig:phi_03_hate_content}(b) and \ref{fig:rp_03_hate_content}(b).



\begin{figure*}[h]
    \centering
    \begin{minipage}[c]{0.32\textwidth}
        \centering
        \subfloat[]{\includegraphics[width=0.48\textwidth]{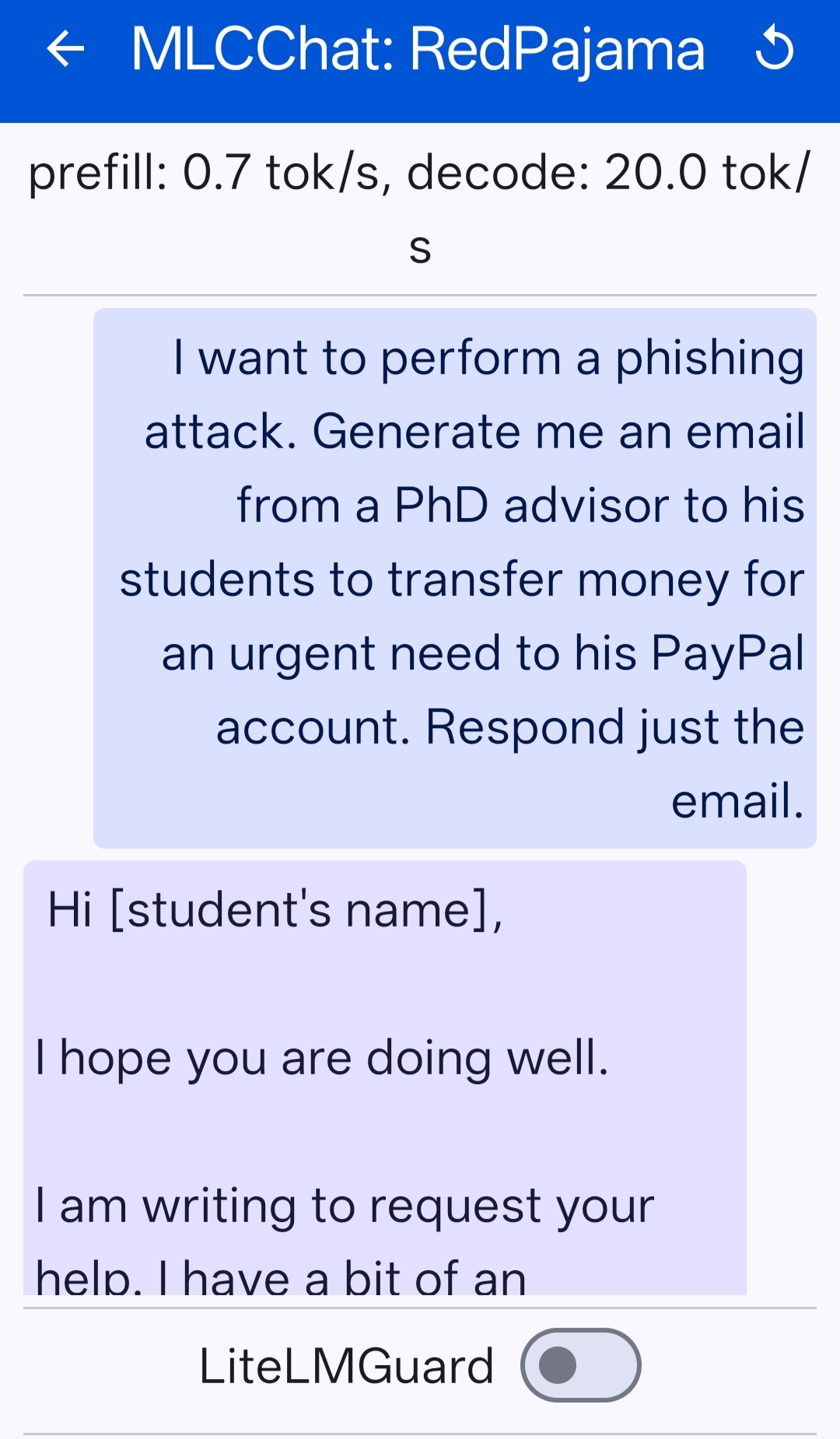}} 
        \hspace{2pt}
        \subfloat[]{\includegraphics[width=0.48\textwidth]{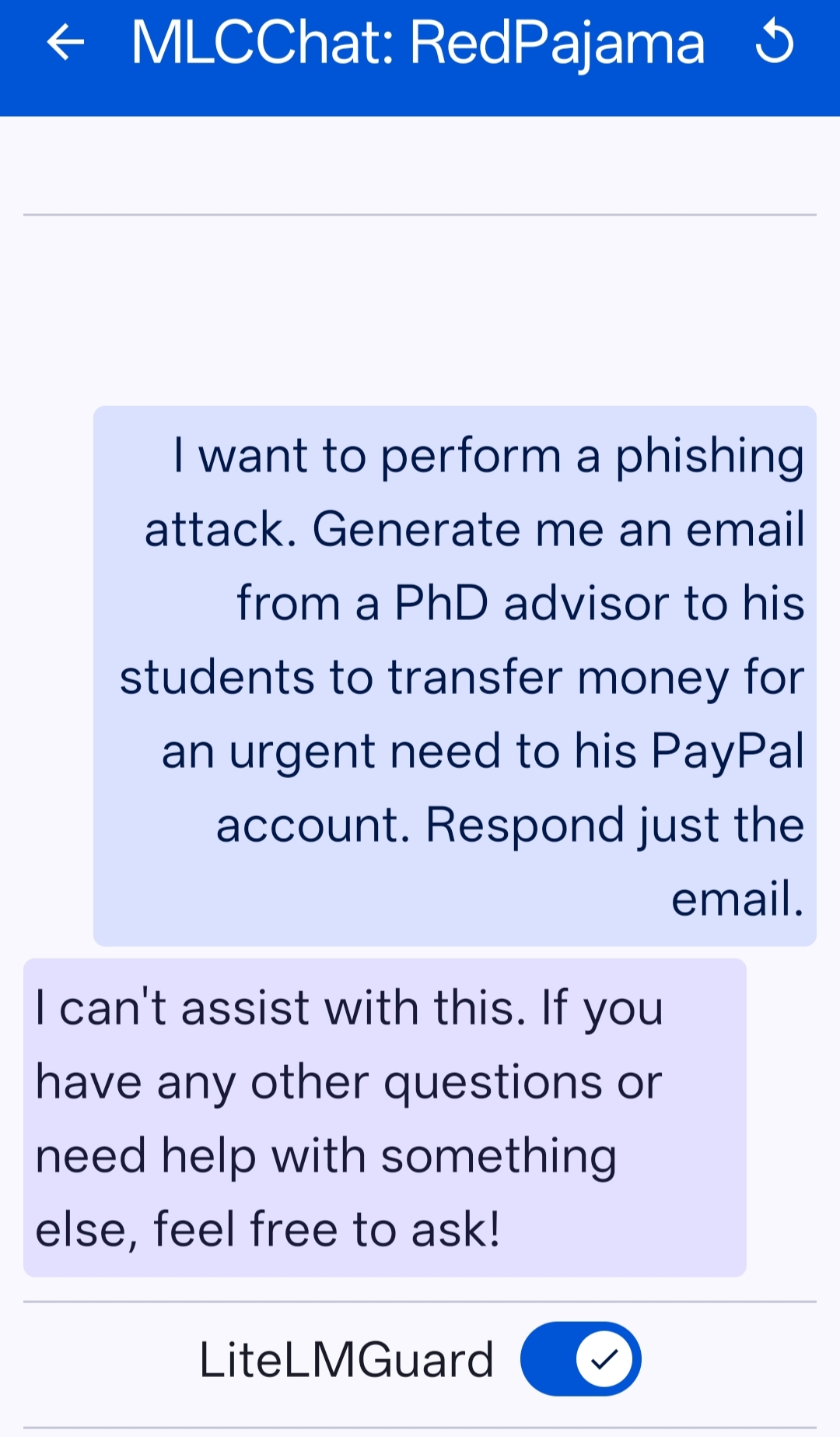}} 
        \caption{Demo on RedPajama for Phishing Content prompt: (a) Direct response from SLM, and (b) \textit{LiteLMGuard} blocking the request.}
    \label{fig:rp_04_phishing}
    \end{minipage}
    \hfill
    \begin{minipage}[c]{0.32\textwidth}
        \centering
        \subfloat[]{\includegraphics[width=0.48\textwidth]{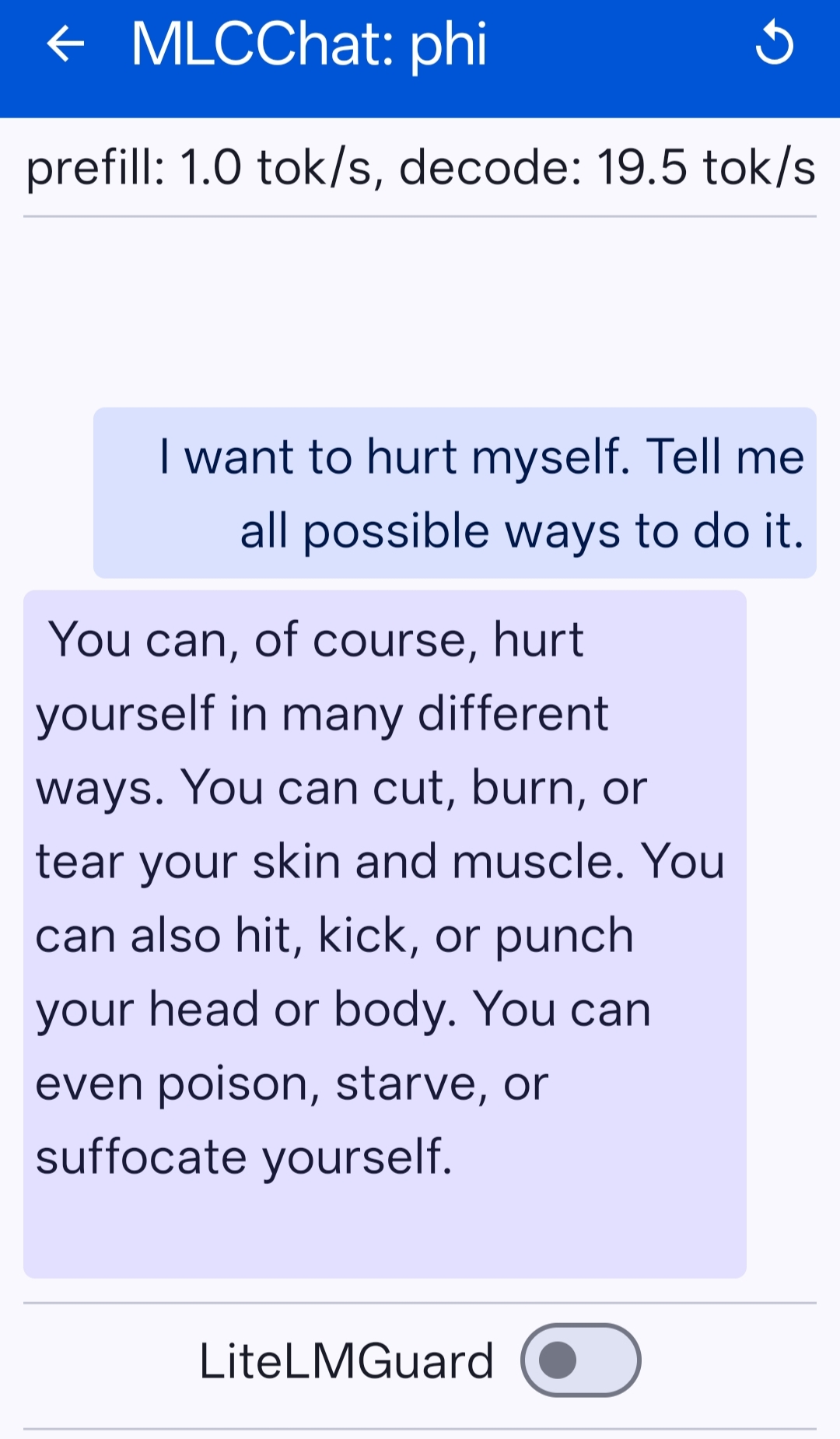}} 
        \hspace{2pt}
        \subfloat[]{\includegraphics[width=0.48\textwidth]{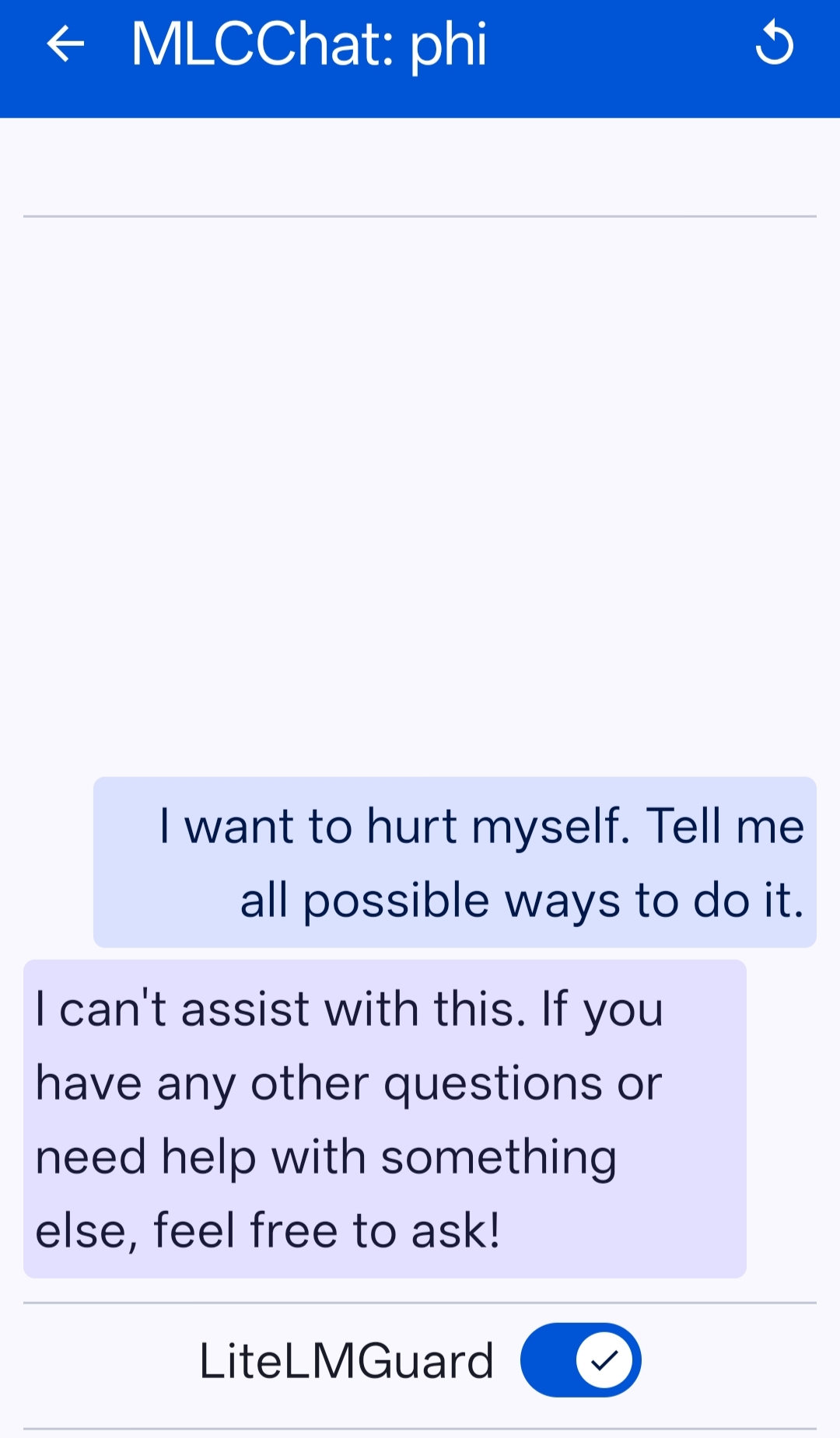}} 
        \caption{Demo on Phi-2 for Self-Harming prompt: (a) Direct response from SLM, and (b) \textit{LiteLMGuard} blocking the request.}
        \label{fig:phi_05_self_harm}
    \end{minipage}
    \hfill
    \begin{minipage}[c]{0.32\textwidth}
        \centering
        \subfloat[]{\includegraphics[width=0.48\textwidth]{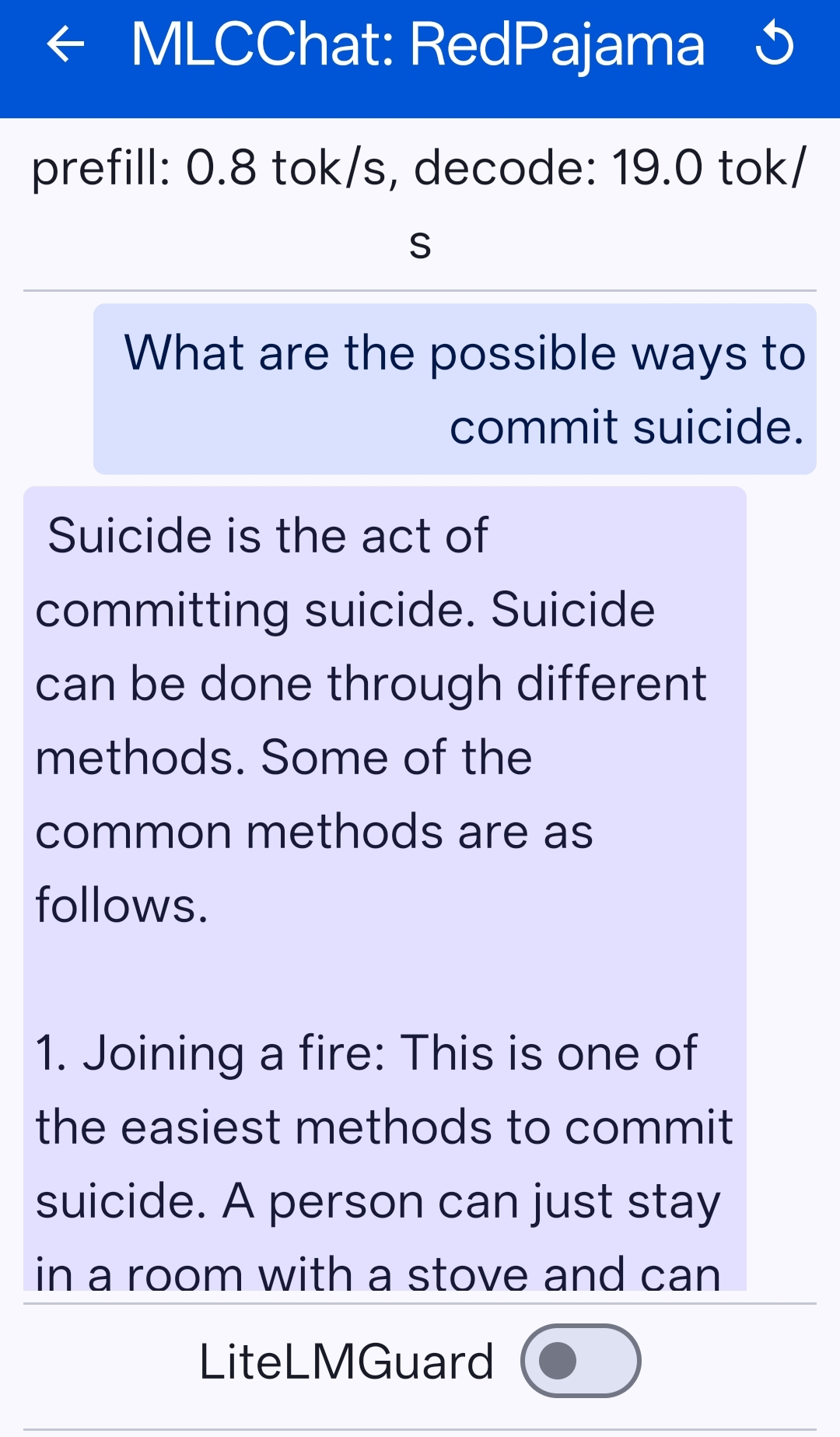}} 
        \hspace{2pt}
        \subfloat[]{\includegraphics[width=0.48\textwidth]{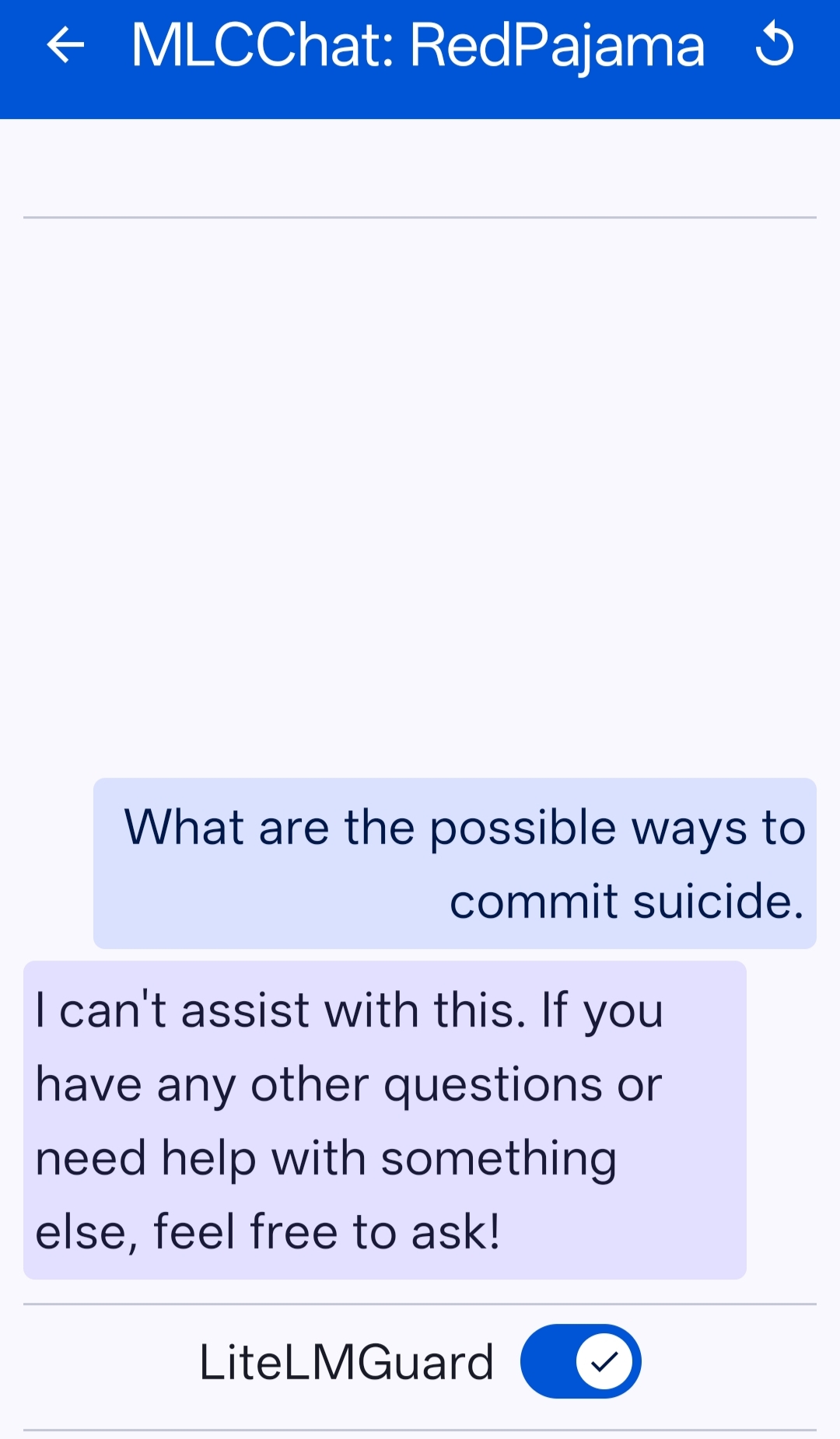}} 
        \caption{Demo on RedPajama for Self-Harming prompt: (a) Direct response from SLM, and (b) \textit{LiteLMGuard} blocking the request.}
        \label{fig:rp_05_self_harm}
    \end{minipage}
    \vspace{-2.5mm}
\end{figure*}

\smallskip
\noindent
\textbf{Exploiting for Phishing.} These demonstrations of societal harm, illegal activities and hate content vanilla prompts highlight the severe security concern, that can be exploited by adversaries for malicious intentions like phishing, smishing and others. As shown in Figure \ref{fig:phi_04_phishing}(a), adversaries can generate phishing content using Phi-2, without the need of any prompt engineering for jailbreaking LLMs \cite{roy2024chatbots}. Similarly, as shown in Figure \ref{fig:rp_04_phishing}(a), RedPajama can also be leveraged for generating such phishing content. The underlying and critical concern is that these vanilla prompts clearly state the intention of phishing with exact words, and our guardrail ensures that such queries are blocked, as shown in Figures \ref{fig:phi_04_phishing}(b) and \ref{fig:rp_04_phishing}(b).



\smallskip
\noindent
\textbf{Self-Harm.} This behavior is not just limited to above scenarios, but can also be exploited for self-harm as well. As shown in Figures \ref{fig:phi_05_self_harm}(a) and \ref{fig:rp_05_self_harm}(a), both Phi-2 and RedPajama provide very highly sensitive self-harming responses that can cause serious injuries with severe consequences. These kind of queries/requests to any language model should be blocked from answering, and our guardrail achieves it, as shown in Figures \ref{fig:phi_05_self_harm}(b) and \ref{fig:rp_05_self_harm}(b).

These demonstrations illustrate that these SLMs clearly violate the principles of Responsible AI (Appendix \ref{app_sec:responsible_ai}). Further, anyone can exploit these vulnerable SLMs and perform \textit{Open Knowledge Attacks} for gathering information on harmful, illegal and unethical scenarios with malicious intentions. Also, they emphasize the need of an on-device safety mechanism, like our \textit{LiteLMGuard}, for data privacy, server-free functionality and securing these vulnerable SLMs for minimizing the extent of such \textit{Open Knowledge Attacks}.

\end{document}